\documentclass{aastex631} 

\def\gtorder{\mathrel{\raise.3ex\hbox{$>$}\mkern-14mu
    \lower0.6ex\hbox{$\sim$}}}
\def\ltorder{\mathrel{\raise.3ex\hbox{$<$}\mkern-14mu
    \lower0.6ex\hbox{$\sim$}}}
    
\usepackage{graphicx}	
\usepackage{amsmath}	
\usepackage{amssymb}	
\usepackage{soul}
\usepackage{newtxtext,newtxmath}

\shorttitle{Galactic Bars at Cosmic Dawn}
\shortauthors{Bi. Da et al.}
\graphicspath{{./}{figures/}}

\begin{document}

\title{Modeling Evolution of Galactic Bars at Cosmic Dawn}

\author{Da Bi}
\affiliation{Department of Physics and Astronomy, University of Kentucky,\\
Lexington, KY 40506-0055, USA}

\author{Isaac Shlosman}
\affiliation{Department of Physics and Astronomy, University of Kentucky,\\
Lexington, KY 40506-0055, USA}
\affiliation{Theoretical Astrophysics, Graduate School of Sciences, Osaka University,\\
Osaka, Japan}


\author{Emilio Romano-D\'{\i}az}
\affiliation{Argelander-Institut f\"ur Astronomie, University of Bonn, \\
Auf dem H\"ugel 71, D-53121 Bonn, Germany}




\begin{abstract}
We study evolution of galactic bars using suite of very high-resolution zoom-in cosmological simulations of galaxies at $z\sim 9-2$. Our models were chosen to lie within similar mass DM halos, log\,($M_{\rm vir}/{\rm M_\odot})\sim 11.65\pm 0.05$, at $z=6$, 4, and 2, in high and low overdensity environments. We apply two galactic wind feedback mechanisms for each model. All galaxies develop sub-kpc stellar bars differing in their properties. We find that (1) The high-$z$ bars form in response to various perturbations: mergers, close flybys, cold accretion inflows along the cosmological filaments, etc.; (2) These bars account for large-mass fraction of galaxies; (3) Bars display large corotation-to-bar-size ratios, and are weaker compared to their low-redshift counterparts, by measuring their Fourier amplitudes, and are very gas-rich; (4) Their pattern speed does not exhibit monotonic decline with time due to braking against DM, as at low $z$; (5) Bar properties, including their stellar population (SFRs and metal enrichment) depend sensitively on prevailing feedback; (6) Finally, we find that bars can weaken substantially during cosmological evolution, becoming weak oval distortions --- hence bars are destroyed and reformed multiple times unlike their low-$z$ counterparts.  In all cases, bars in our simulations have been triggered by interactions. In summary, stellar bars appear to be not only contemporary phenomenon, but based on increased frequency of mergers, flybys and the strength of cold accretion flows at high $z$, we expect them to be ubiquitous at redshifts $\gtorder 2$ --- the epoch of rapid galaxy growth and larger stellar dispersion velocities.  

\end{abstract}



\section{Introduction} \label{sec:intro}

A large fraction of local galaxies, $\sim 2/3$ of them at least,  host stellar bars, both strong and weak \citep[e.g.,][]{sell93,marti95,kna00,gros04,mene07,erwin18}. Bars play the prime role in redistributing the angular momentum in galaxies, primarily between the disks and their parent dark matter (DM) halos, in tandem with galaxy interactions and mergers \citep[e.g.,][]{lynd72,trem84a,wein85,deba00,atha03,marti06}. Bars are also responsible for mass redistribution in galaxies, especially of the gas. Thus they act as a prime internal driver of the galaxy evolution.  

The origin of bars is not known precisely --- they can form spontaneously by a global gravitational instability in axisymmetric stellar disks \citep[e.g.,][]{hohl71,sell80,toomre81}, can be tidally induced by close flybys and mergers \citep[e.g.,][]{ger90,bere04,marti17}, or form in response to the dark matter (DM) halo shape and DM substructure \citep{hel07,roma08}. In short, bars can form as a result of the bar instability, or in response to a finite perturbation.

Galactic bars come in variety of sizes. Some can be compared in size to their parent galactic disks, and some are much smaller. The reason for this diversity is probably related to the combination of factors, such as the mass distribution in galaxies, the stellar and gas dispersion velocities in the disk, the efficiency of angular momentum redistribution in the disk-halo systems, stage in their evolution, etc.

Here we focus on cosmological evolution of bars at redshifts $z\gtorder 2$, and follow them by means of very high-resolution zoom-in numerical simulations. We analyze the basic parameters of these bars, relate their evolution to both internal and external factors, and compare them with low-$z$ bars.

Our goal to is to compare the evolution of barred galaxies at high redshifts with those at low redshifts, published in the literature. We also point out a number of corollaries that follow the different evolution. We address this issue by comparing bars which form and evolve in galaxies that are embedded in similar mass dark matter (DM) halos at different redshifts. We choose halos which have virial masses log\,($M_{\rm vir}/{\rm M_\odot})\sim 11.65\pm 0.05$, at the final redshifts  of $z_{\rm f}=$ 6, 4 and 2. This DM halo mass is a factor of 2.5 smaller than that of the Milky Way halo obtained from GAIA \citep[e.g.,][]{posti19}. Such halos are mostly typical at $z=2$ ($1\sigma$), more scarce at $z=4$ ($2\sigma$), and rare at $z=6$ ($2.5\sigma$) \citep[e.g.,][where $\sigma^2$ is the variance of the linear density field.]{reed03}. Our choice of halos includes the specific environment as measured by the local overdensity, as well as a similar spin (more details given in \citet[][hereafter Paper\,I]{bi21} and in Table\,\ref{tab:DMsim}.

Our motivation for this particular DM halo masses is based on the following factors: (1) avoiding low-mass galaxies which cannot form sustainable galactic disks, and avoiding the high mass halos which would be exceedingly rare at these redshifts; (2) As similar mass halos at different redshifts grow at different rates, this effect is expected to propagate down to galactic scales, as shown and analyzed in Paper\,I; (3) Focusing on similar mass halos at different redshifts allows us to single out two processes, the effect of environment and that of the stellar feedback; (4) Moreover, such halos are expected to host galaxies with stellar masses in excess of $10^{10}\,{\rm M_\odot}$, being most hospitable to developing stellar bars, based on observations and modeling in the contemporary universe \citep[e.g.,][]{gad09}; (5) Finally and most importantly, these halos contain $L*$ galaxies at $z=0$, so the comparison of $z\gtorder 2$ galaxies with the present day galaxies is the most optimal one. 

Observational studies of the bar fraction evolution with redshift have been attempted so far only for $z\ltorder 1$. Disk galaxies show that the resolved bar fraction (in galaxies with a single bar) is constant within a factor of two for $z\sim 0-1$ \citep{jog04}. On the other hand, \citet{sheth08} claimed that the bar fraction decreases with increasing redshift. Both studies can be reconciled if only large size bars,  which can be unambiguously detected up to $z\sim 1$, are counted. However, small size bars can be overlooked even at $z\sim 0$, especially when searched in the optical band. For higher redshifts, the properties of bars and their abundance are completely unknown.

Numerical simulations of stellar bars require sufficient spatial and mass resolutions to adequately follow their nonlinear dynamics. Consequently, most detailed and reliable simulations of barred galaxies are those of isolated objects which allow to resolve the resonant interactions between bars, surrounding disks, and their parent DM halos \citep[e.g.,][]{atha02,atha03,marti06}. The necessary number of particles to follow this evolution properly has been claimed to lie in the range of $\gtorder 10^8$ disk stars and DM particles \citep{wein07a,wein07b}, although this requirement has been shown to be excessive, and $\gtorder 10^6$ stellar/DM particles can suffice \citep{dubi09}.

Spontaneously forming bars in isolated disk-halo systems grow their amplitude exponentially until saturation. Reaching maximal strength, they experience the so-called vertical buckling instability which weakens and shortens them substantially \citep[e.g.,][]{comb81,comb90,raha91,pfenn91,friedli93,marti04,marti06,coll20}. 

The subsequent evolution depends on the parent DM halo dimensionless spin, e.g., defined as $\uplambda$, \citep[e.g.,][]{bull01},
\begin{equation}
\uplambda = J/J_{\rm max},    
\end{equation}
where $J$ and $J_{\rm max}$ are the DM halo angular momentum and maximal (Keplerian) angular momentum, respectively. Inside halos with a low spin, $\uplambda \ltorder 0.03$, bars resume their growth after buckling, while for halos with a higher spin, the bar growth becomes progressively weaker, until there is no growth at all \citep{long14,coll18}. For $\uplambda\gtorder 0.05$, bars essentially dissolve after buckling, leaving behind only a weak oval distortion.

Cosmological simulations of stellar bars have been limited, and suffered from insufficient spatial and mass resolution. Over the last decade, in the zoom-in cosmological simulations, the mass per baryonic and DM particles went below $10^6\,{\rm M_\odot}$, and the number of baryonic and DM particles has approached $\sim 10^{5-6}$ per galaxy-halo system \citep[e.g.,][]{roma08,scan12,oka15,zana19}.

Cosmological simulations based on large computational boxes, e.g., comoving $50 - 100$\,Mpc, have been also analyzed aiming at statistics of the barred galaxies. The Illustris-1 \citep{vogel14}, IllustrisTNG100-1 \citep{nel18}, TNG50 \citep{nel19} and EAGLE \citep{schaye15} simulations provided a rich ground for data mining for redshifts $z\ltorder 2$, with interesting conclusions on the bar evolution \citep[e.g.,][]{algor17,pes19,rosa20,rosa21,zhao20}. We refer to these simulations in section\,\ref{sec:discussion}, but note here that the mass and spatial resoloution of these simulations is substantially lower than in our runs, with the exception of TNG50 \citep{rosa20} which has only about 3 times lower mass resolution.

The zoom-in simulations of stellar bar evolution aimed at {\it relatively} isolated galaxies, in order to mimic the Milky Way galaxy \citep[e.g.,][]{scan12,oka15,zana19,bla20}. \citet{roma08}, using zoom-in cosmological simulation, followed the bar evolution from $z\sim 10$ to $z=0$, including an intensive merger activity at $z\sim 9 - 3$. This simulation has captured some key points of the bar evolution: the tidal triggering of the bar and the variable strength and pattern speed of the bar due to galaxy mergers and interactions. Most importantly, it showed that the bar amplitude and pattern speed can increase and decrease over prolonged time periods.

The physical conditions in our current simulations are not constrained by the quiet evolution of the halos and include mergers, flybys, cold accretion flows and more. The parent DM halos of modelled galaxies are situated in low and high overdensity environments (see section\,\ref{sec:sims} for definition). In section\,\ref{sec:discussion}, we compare our results obtained for higher redshifts, $z\gtorder 2$, with those in the above simulations. We anticipate that the low redshift evolution is more compatible with the simulations of isolated galaxies or those outside the clusters of galaxies.

An important question can be asked, to what extent the evolution of galactic bars in isolated galaxy models can adequately describe those in the full cosmological context. The possible answer can depend on the redshift under consideration. It is especially relevant for galaxy evolution at $z > 1$, when cold accretion fuels the galaxy growth in tandem with mergers and close flybys. Galactic disks are expected to grow with time, roughly up to $z\sim 1$, when the growth slows down. This saturation of the disk growth is most probably associated with decrease in the frequency of mergers and cold accretion flows. So we do expect that low redshift galaxies have prolonged periods of quiescent growth, which has a direct consequence for stellar bar evolution. 

We focus on the population of galactic bars at high redshifts, $z\gtorder 2$, and compare them with the bar properties at low $z$, in preparation with the forthcoming observations. The high--redshift galaxy evolution proceeds in a denser and more violent universe compared to the low redshift evolution, $z\ltorder 1$, which is more compatible with simulations of isolated galaxies. As the bar evolution is closely related to that of their host galaxies and even their DM halos, we analyze them as well in this context, although the full details of parent galaxies and their halos evolution are provided in Paper\,I. 

The structure of this paper is as following. We briefly discuss the bar properties in the local universe in section\,\ref{sec:locals}, describe our numerical methods and initial conditions in section\,\ref{sec:numerics}, and present our results in section\,\ref{sec:results}. Section\,\ref{sec:discussion} deals with corollaries of high-redshift bar evolution, followed by conclusions.

\section{Stellar bar properties in the local universe} \label{sec:locals}

Observational and theoretical properties of local bars have been extensively reviewed in the literature \citep[e.g.,][]{sell93,shlo99,kor04,atha13,shlo13}, and we list them only briefly in this section. 

Bars typically extend to nearly their corotation radius (CR), where the bar pattern speed, $\Omega_{\rm bar}$, is equal to the circular stellar frequency in the disk, $\Omega$, and never beyond it, because the stellar orbits change their orientation by $\pi/2$ at every resonance. Inside the CR, the stellar orbits are generally aligned with the bar, forming and supporting its density distribution. Outside the CR the stellar orbits are oriented normal to the bar and hence cannot support its shape. The ratio of the CR-to-bar radius has been found to lie within the range of $R_{\rm CR}/R_{\rm bar} = 1.2\pm 0.2$ \citep{atha92}. The narrow range of this ratio is based on the observed shape of the offset dust lanes, associated with shocks within the bars, and it is confirmed by observations of local bars \cite[e.g.,][]{cuomo19,guo19},  when the bar pattern speed has been determined using the Tremaine-Weinberg method \citep{trem84b}. However, this ratio was found to be violated during the buckling instability \citep{marti06} and in faster spinning halos \citep{coll18}.  The latter effect appears to be strongest for the prolate halos. \citet{marti17} have detected that bars triggered by a flyby approximated by an impulse approximation show a more extended time when $R_{\rm CR}/R_{\rm bar}$ stays outside the above narrow range.

Bars that are well short of their CR radius are called slow bars, and those that extend (approximately) to CR are named fast bars. Stellar bars are known to brake against the DM and slow down with time, but also grow in length during this process by capturing disk stars. Hence the slow bars have difficulty to grow. Is this a difficulty limited to numerical simulations only?

A separate class of bars extending to the Inner Lindblad resonance, the ILR\footnote{The ILR is defined as the radius of intersection of frequencies, namely, the bar pattern speed, $\Omega_{\rm bar}$, and the combination of circular frequency $\Omega$ and the epicyclic frequency, $\kappa$, i.e., $\Omega_{\rm bar} = \Omega - \kappa/2$.}, proposed by \citet{lynd79}, have not been identified so far. The only exception are the nuclear bars in double barred galaxies whose size does not correlate with the galaxy size and extend to the ILR of the large bar \citep{laine02}.

In the local universe, the median observed bar sizes lie in the range of $3-4$\,kpc \citep[e.g.,][]{erwin05,mene07,mari07,cors11,ague15,font17}. However, the size distribution is quite broad, from a fraction of kpc to about 15\,kpc and depends on the observational band. From the theoretical point of view, this size scatter is understood to come from a number of factors (section\,\ref{sec:intro}), the most obvious reasons are the baryonic and DM mass distributions, the efficiency of angular momentum redistribution in the disc-halo systems, and stellar dispersion velocities in the parent disks. The size distributions in the $B$ and $H$-bands peak around the median, but decrease more sharply towards the small sizes in the $B$-band compared to the $H$-band. In the extreme case, galaxies that have been classified as unbarred in optical, display a strong bar in the $K$-band \citep[e.g.,][]{scov88,teles88,thron89,kna95}.  

The vertical extension of stellar bars is affected by the buckling instability, leading to the appearance of characteristic boxy/peanut shape bulges which are thicker than the underlying galactic disks \citep[e.g.,][]{bure99,merri99}. Observationally about 50\% of edge-on disks display these bulges \cite[e.g.,][]{luti00}. Strong boxy/peanut shape bulges have been reproduced in numerical simulations of collisionless disks. Addition of gas component was found to weaken the buckling instability and smooth the bulge shape, which was attributed to the destabilization of stellar orbits by the gas accumulation in the central regions \citep{bere98}.  

Galactic bars in the local galaxies are almost never observed to be dominated by gas. Local bars are either largely devoid of gas or live in galaxies with the gas fraction of less than 50\%, smaller than in the unbarred galaxies \citep[e.g.,][]{davo04}. The simplest explanation is that bars channel the disk gas inwards, where the gas can be converted into H$_2$ and form stars. The only exclusion can be traced to the inner bars of double barred galaxies, where the gaseous component can be significant and even dominating \citep[e.g.,][see also review by \citet{shlo05}]{shlo89,friedli93,maio00,engl04}.

Stellar populations of local bars are typically dominated by older disk stars and this explains why bars are more easily detectable in the $K$-band. Strong bars show low star formation rate, mostly concentrated in their centers and on their ends. This is related to the strong shear affecting the giant molecular clouds. On the other hand, weak bars do exhibit star formation. A good local example of a mild bar experiencing the star formation along the offset shocks is M100 \citep[e.g.,][]{kna95}.

Numerical simulations of bars in isolated disk-halo systems emphasize that the bar pattern speeds decrease as a result of angular momentum loss to the outer disk and especially to the DM halos \citep[e.g.,][]{sell80,atha02,atha03,marti06,villa09}. Braking against the DM and stars has a two-fold nature. It is facilitated mainly by resonances, but a non-resonant braking should not be discarded as well. 

Loss of the angular momentum by the bar leads to the change in the bar pattern speed (i.e., its tumbling) and to the change of the internal angular momentum of the bar (i.e., internal circulation). The former leads to the bar slowdown, the latter --- to the increase in the bar strength. Overall, except for some brief moments, the bar region is losing its angular momentum monotonically. In some cases this can create controversies --- locally observed bars tumble faster than predicted by numerical simulations of gas-poor isolated galaxies.  

The bar fraction in the local universe does not correlate with the environment, e.g., with local density \citep[e.g.,][]{ague09}.  Moreover. \citet{mari09} found that the cluster environment does not strongly affect the bar fraction. 

\section{Numerical modeling} \label{sec:numerics}
\begin{table*}
\resizebox{1.0\textwidth}{!}{%
\label{tab:table1}
\centering
\begin{tabular}{ccccccccccccc} 
\hline
             &&&   DM Halo Properties in baryonic simulations &&&&  && Galaxy Properies \\
\hline
 $z_{\rm f}$ & Model & ${\rm log}M_{\rm vir}\,{\rm M_\odot}$ & $R_{\rm vir}$\,kpc & $R_{\rm vir}$\,kpc & $\uplambda$ & $\updelta$ & $M_*\,{\rm M_\odot}$ & $R_{\rm gal}$\,kpc & $R_{\rm gal}$\,kpc & $f_{\rm gas}$ & B/T & Feedback \\
          & Name &  & comoving & physical & &   &  & comoving & physical &  & kinematic &\\
\hline
\hline
    6     & Z6HCW & 11.6   & 252 & 36 & 0.02  & 3.04 &$2.54\times 10^{10}$ & 9.8  & 1.4 & 0.50 & 0.57 & CW  \\
    6     & Z6HVW & 11.6   & 257 & 37 & 0.02  & 3.04 &$0.59\times 10^{10}$ & 9.8  & 1.4 & 0.87 & 0.53 & VW  \\
\hline   
    6     & Z6LCW & 11.6   & 251 & 36 & 0.02  & 1.60 &$1.97\times 10^{10}$ & 12.1 & 1.7 & 0.52 & 0.57 & CW  \\   
    6     & Z6LVW & 11.6   & 249 & 36 & 0.02  & 1.60 &$0.48\times 10^{10}$ & 8.6  & 1.2 & 0.87 & 0.63 & VW  \\   
\hline  
\hline   
    4     & Z4HCW & 11.6   & 271 & 54 & 0.06  & 3.00 &$3.21\times 10^{10}$ & 7.9  & 1.6 & 0.25 & 0.59 & CW  \\   
    4     & Z4HVW & 11.6   & 254 & 51 & 0.06  & 3.00 &$0.72\times 10^{10}$ & 9.5  & 1.9 & 0.85 & 0.50 & VW  \\  
\hline
    4     & Z4LCW & 11.6   & 258 & 52 & 0.02  & 1.33 &$3.30\times 10^{10}$ & 6.8  & 1.4 & 0.38 & 0.75 & CW  \\
    4     & Z4LVW & 11.6   & 267 & 53 & 0.02  & 1.33 &$1.05\times 10^{10}$ & 6.9  & 1.4 & 0.86 & 0.67 & VW  \\
\hline
\hline   
    2     & Z2HCW & 11.7  & 293  & 98 & 0.02  & 2.80 &$6.61\times 10^{10}$ & 11.2  & 3.7 & 0.28 & 0.80 & CW  \\   
    2     & Z2HVW & 11.7  & 293  & 98 & 0.02  & 2.80 &$3.10\times 10^{10}$ & 13.4  & 4.5 & 0.77 & 0.80 & VW  \\ 
\hline
    2     & Z2LCW & 11.7  & 283  & 94 & 0.02  & 1.47 &$5.31\times 10^{10}$ & 10.6  & 3.5 & 0.27 & 0.62 & CW  \\
    2     & Z2LVW & 11.7  & 261  & 87 & 0.02  & 1.47 &$1.52\times 10^{10}$ & 7.7   & 2.6 & 0.76 & 0.65 & VW  \\
\hline   
\hline
\end{tabular}
}
\caption{Simulation suite (all values are given at the final redshift $z_{\rm f}$); the model (see section\,\ref{sec:winds}); $M_{\rm vir}$ is the virial mass of DM halo in baryonic simulations; $R_{\rm vir}$ is the halo virial radius in comoving and physical coordinates; $\uplambda$ is the DM halo spin; $\updelta$ is the local overdensity; $M_*$ is the stellar mass of the central galaxy; $R_{\rm gal}$ is the galaxy radius in comoving and physical coordinates; $f_{\rm gas}$ is the gas fraction of the central galaxy; B/T --- bulge-to-total stellar mass, based on kinematic decomposition (see paper\,I); CW and VW are the galactic wind feedback, Constant Wind (CW) and Variable Wind (VW).  }
\label{tab:DMsim}
\end{table*}

\subsection{Cosmological zoom-in simulations and initial conditions}
\label{sec:sims}

In this work we analyze our simulations which used the hybrid $N$-body/hydro code \textsc{gizmo} \citep{hopk17}. We run a suite of zoom-in cosmological simulations within a box of 74\,Mpc comoving coordinates. We assume the $\Lambda$CDM cosmology with parameters $\Omega_{\rm m} = 0.308$, $\Omega_\Lambda = 0.692$ and $\Omega_{\rm baryonic} = 0.048$, $\sigma_8=0.82$, and $n_8=0.97$ \citep{planck16}. The Hubble constant is taken as $h = 0.678$ in units of $100\,{\rm km\,s^{-1}\,Mpc^{-1}}$. The simulations have been run from $z = 99$ to $z_{\rm f} = 6$, 4, and 2. Additional details of the models are given in Table\,1  and in Paper\,I. 

We invoked the meshless finite mass (MFM) hydrosolver and an adaptive gravitational softening for the gas. The \citet{spri03} multiphase interstellar matter (ISM) algorithm has been used, and the feedback includes the SN\,II, and two models of galactic winds (see section\,\ref{sec:winds}). Feedback by SN\,Ia appears to be unimportant for $z\gtorder 2$ \citep[e.g.,][]{child14}, and therefore was not included.  The gas cooling includes the Compton, free-free, collisional and metals. Ionization and recombination processes have been accounted for as well. Simulations include the redshift-dependent cosmic UV background \citep{Faucher09}. The density threshold for star formation (SF) was set to $n^{\rm SF}_{\rm crit}=4\,{\rm cm^{-3}}$.   

\newpage
\subsubsection{Defining parent DM halos}
\label{sec:halos}

The parent box and the individual zoom-in simulations were created using \textsc{music} \citep{hahn11}. As a first step, we have generated a comoving box of 74\,Mpc with $1024^3$ DM-only particles, and run it until redshift $z=2$. Applying the group finder (see below), the DM halos has been selected at the final redshifts of $z_{\rm f} = $ 6, 4, and 2, that have the prescribed DM masses of log\,($M_{\rm vir}/{\rm M_\odot})\sim 11.6$, dimensionless halo spin $\uplambda$, and the local overdensity $\updelta$ (see below for the definitions of $M_{\rm vir}$ and $\updelta$). 

We calculated halos by using the \textsc{rockstar} group finder \citep{behr13}, accounting only for the bound DM particles (Table\,\ref{tab:DMsim}). Using the \textsc{rockstar} halo catalogs, we have constructed the merger trees using \textsc{consistent-trees} algorithm \citep{behr12}. The halo virial radius  and the virial mass, $R_{\rm vir}$ and $M_{\rm vir}$, are defined by $R_{200}$ and $M_{200}$ \citep[e.g.,][]{nfw96}. $R_{200}$ is the radius inside which the mean interior density is 200 times the critical density of the universe at that time.

Next, we have carved out a sphere encompassing all particles within the volume of $\sim 3R_{\rm vir}$ to avoid contamination of the high resolution region by the massive particles. These particles have been traced to their initial conditions, in order to create a mask for \textsc{music}, where the resolution has been increased by 5 levels, i.e., from $2^7$ to $2^{12}$. Hence, we have 5 levels of refinement and analyze only the highest level --- our mass and spatial resolution given below refer only to this level.

The overdensity environment $\updelta$ has been defined in the parent run (i.e., full box run) by creating 1.5\,Mpc grids centered on DM halos and calculating the average DM densities inside these grids. The ratio between these densities and the average density of the universe provide $\updelta$. The DM halos have been chosen for low and high overdensities, $\updelta\sim 1$ and $\updelta\sim 3$, respectively.

\subsubsection{Defining central galaxies}
\label{sec:galaxies}

In order to identify galaxies, we have used the group finder \textsc{hop} \citep{eise98}, in terms of the boundary baryonic density threshold of $10^{-2}n_{\rm crit}^{\rm SF}$, ensuring that both the host starforming and non-starforming gas are roughly bound to the galaxy \citep{roma14}. \textsc{HOP} has been used in order not to impose a particular geometry on galaxies. Our galaxies morphological and kinematic parameters are given in Table\,\ref{tab:DMsim} (see also  Paper\,I). 

We have abbreviated the galaxy names in terms of their simulated properties in the following way: the final redshift $z_{\rm f}$, the overdensity $\updelta$ (high or low), and the stellar feedback (CW or VW, see section\,\ref{sec:winds} for definition of the feedback). For example, the galaxy Z4HCW corresponds to $z_{\rm f}=4$, high overdensity, and with the CW feedback scheme.

In our simulations, the mass per particle is $3.5\times 10^4\,{\rm M_\odot}$ (for gas and stars) and $2.3\times 10^5\,{\rm M_\odot}$ (for DM). The minimal adaptive gravitational softening is 74\,pc for the gas in comoving coordinates. The softening for stars is 74\,pc and 118\,pc for DM (in comoving coordinates). This means, for example, that at the final redshifts, $z_{\rm f}=6$, 4, and 2, the softening for the stars in physical coordinates is 10.5\,pc, 14.7\,pc, and 24.6\,pc, respectively.  The effective number of baryonic particles in our simulations is $2\times 4096^3$. 

For comparison, the Illustris-1, IllustrisTNG100-1 and TNG50 simulations \citep{vogel14,nel18,nel19} have the DM mass per particle $6.3\times 10^6\,{\rm M_\odot}$, $7.5\times 10^6\,{\rm M_\odot}$,  and $4.5\times 10^5\,{\rm M_\odot}$, respectively. The gas and stellar mass per particle is $1.6 \times 10^6\,{\rm M_\odot}$, $1.4\times 10^6\,{\rm M_\odot}$ and $8.5\times 10^4\,{\rm M_\odot}$. The EAGLE simulations \citep{schaye15} have the mass resolution of $9.7\times 10^6\,{\rm M_\odot}$ for DM and  $1.8\times 10^6\,{\rm M_\odot}$ for the gas. The gravitational softening in Illustris-1, TNG100-1 and TNG50 is 1.4\,kpc, 0.74\,kpc and 0.29\,kpc for DM,  and 0.74\,kpc, 0.19\,kpc and 0.074\,kpc for the gas, respectively. In EAGLE simulations, the resolution used for the bar evolution study by \citet{algor17} is 0.7\,kpc at $z=2.8$, although EAGLE has run also a higher resolution of 0.35\,kpc. This softening is depending on redshift for $z > 1$. Hence, with the exception of the TNG50 simulations, both the mass and spatial resolutions are substantially reduced in comparison to those in the present work, up to two orders of magnitude. The TNG50 is compatible to our simulations within a factor of 2--3.

The mass and spatial resolutions are crucial to follow the resonances which are associated with bars, especially of a smaller size bars. The spatial resolution of bars lies in that a sufficient number of softening lengths must be included in its size, e.g., $\gtorder 6$. The number of stars in a stellar disk should be $\gtorder 10^{5-6}$. Bars which are poorly resolved spatially or with a number of particles will not account for an efficient angular momentum transfer, for accurate internal evolution and additional effects \citep{dubi09}. Our analysis of bar evolution is limited largely to time periods when these conditions have been satisfied.
 
\subsection{Galactic Winds}
\label{sec:winds}

We have used two models for galactic winds, namely, the Constant Wind \citep{spri03} and the Variable Wind \citep{oppe06} (hereafter abbreviated as CW and VW, respectively). The wind models have been implemented by decoupling the gas particles from the hydrodynamical forces, thus the wind particles move ballistically. The decoupling time period is the shortest between $10^6$\,yrs and the time it takes for the particles to reach the background gas density which has decreased by a factor of 10.  

For the CW model, both the wind velocity, $v_{\rm w}$, and the mass loading factor, $\beta_{\rm w}\equiv \dot M_{\rm w}/\dot M_{\rm SF} = 2$, have been taken constant, where $\dot M_{\rm w}$ is the mass loss by the wind, and $\dot M_{\rm SF}$ is the (mass) star formation rate (SFR). The wind velocity for the CW is taken $v_{\rm w} = 484\,{\rm km\,s^{-1}}$.   The wind orientation has been assumed isotropic. 

For the VW model, the launching velocity scales with halo/galaxy overall properties. The wind velocity has been assigned by the physical escape velocity of the host halo. Mass loading factor has been calculated assuming the total wind energy is given by energy-driven wind and momentum-driven wind.  

The goal of introducing two feedback mechanisms is to compare the galaxy evolution, including the galactic bar properties, applying a strong and a much weaker feedback. We have measured the ratio of kinetic energies of CW and VW, $E_{\rm CW}/E_{\rm VW}$, and find that this ratio is always less then unity. Typically it ranges between 0.1 and 0.01, sometimes diving below these values. Generally, the ratio varies by a factor up to $10^3$. The expected result is that the SFR will be smaller in the VW models, which indeed is the case. Additional corollary of this evolution is that the gas fraction in VW models is larger than in CW ones. 

\subsection{Determining strength, pattern speed, size and mass of stellar bars}
\label{sec:bars}

\begin{figure}
\center 
	\includegraphics[width=0.7\textwidth]{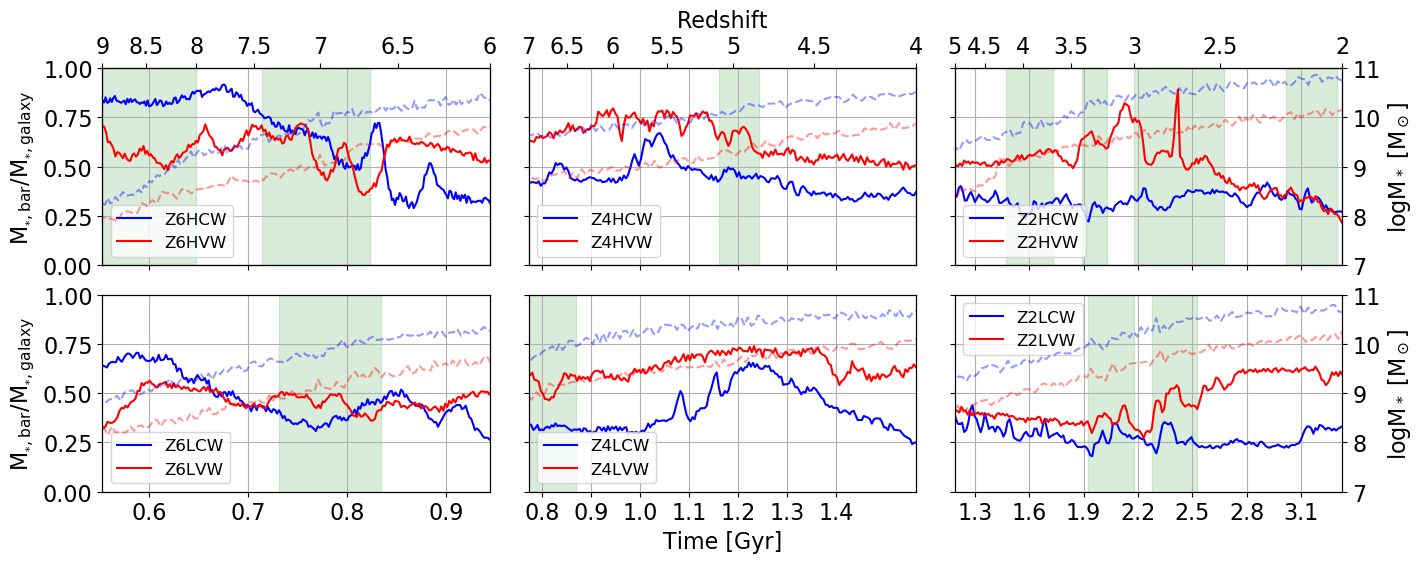}
    \caption{Evolution of stellar bar-to-galaxy mass fraction, $M_{\rm *,bar}/M_{\rm *,galaxy}$ (left y-axes) and galaxy stellar masses, $M_*$ (right y-axes).  The VW models are given by the red solid lines, and CW models by the blue solid lines. The galaxy stellar mass evolution are given by dashed lines. The green colored background refers to major merger events. Note that the minor and intermediate mergers, and close flybys, as well as periods of massive cold accretion inflows can have equally strong effect on the galaxy properties, but are not marked for clarity. }
    \label{fig:starbarmassfrac}
    \end{figure}

 \begin{figure}
\center 
	\includegraphics[width=0.7\textwidth]{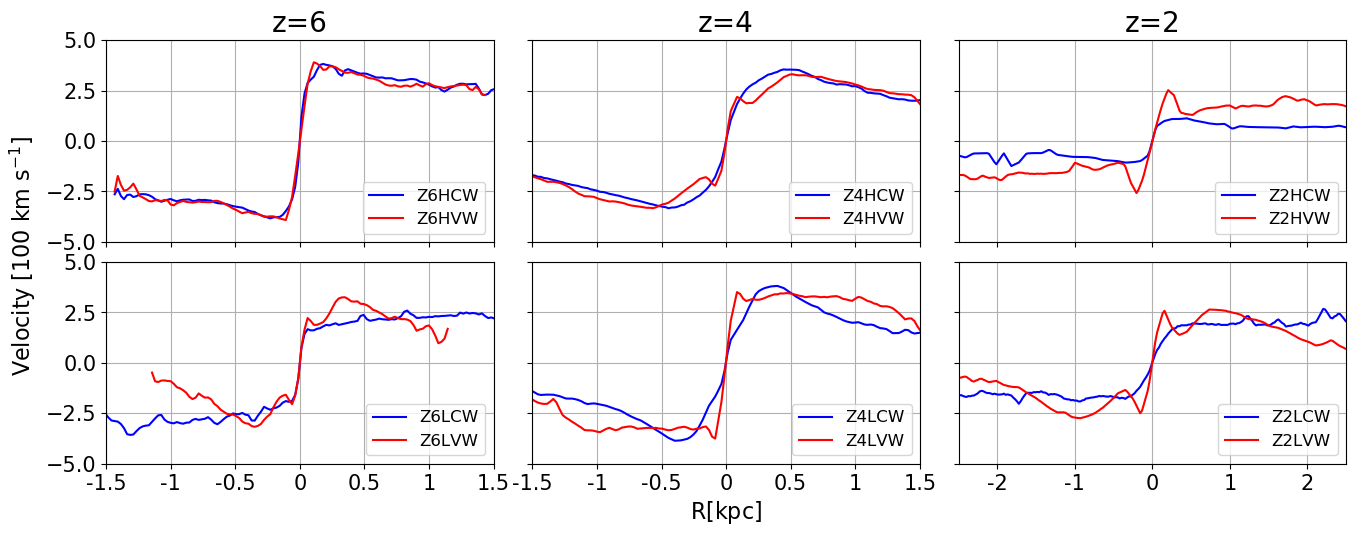}
    \caption{Rotation curves of sample galaxies at the final redshifts, $z_{\rm f} = 6, 4$, and 2, in physical coordinates. The CW models (blue lines) display very similar rotation curves to those of VW models, despite that the later have larger gas fractions. Note, that the top row here and elsewhere in Figures represents galaxies in the high overdensity regions, while the bottom row displays galaxies in the low overdensity regions.}
    \label{fig:galrot}
    \end{figure} 

We analyze the bar properties, such as the bar strength, pattern speed, size and mass evolution using the Fourier components for the surface density for the $m=2$ Fourier mode. This is performed for face-on disks. We obtain the disk plane by calculating the stellar angular momentum vector of stars within the radius containing 90\% of the total stellar content in a galaxy, using its definition in section\,\ref{sec:galaxies}. The Fourier amplitude for the $m$ mode is $A_{\rm m}=\sqrt {a_{\rm m}^2 + b_{\rm m}^2}$, where

\begin{equation}
    a_{\rm m}(r) = \frac{1}{\pi} \int_{0}^{2\pi} \Sigma(r, \theta) \, \cos(m\theta) \, \mathrm{d} \theta, \; \; \; m=0, 1, 2, \ldots
\end{equation}
\begin{equation}
    b_{\rm m}(r) = \frac{1}{\pi} \int_{0}^{2\pi} \Sigma(r, \theta) \, \sin(m\theta) \, \mathrm{d} \theta, \; \; \; m=1, 2, \ldots
\end{equation}
and $\Sigma(r, \theta)$ is the surface stellar density. To quantify the bar strength $A_2$ amplitude, normalized by the $m=0$ amplitude, we use 

\begin{equation}
    \frac{A_2}{A_0}=\frac{ \int_{0}^{R_{\mathrm{bar}}} \sqrt {a_{2}^2 + b_{2}^2} \, \mathrm{d} r}{\int_{0}^{R_{\mathrm{bar}}} a_0 \, dr }.
    \label{eq:barA2}
\end{equation}
where the upper limit, $R_{\mathrm{bar}}$, is the bar radius at a given time, and $A_0$ is the Fourier coefficient for the monopole term. We consider the $m=2$ Fourier mode representing a stellar bar for $A_2/A_0\gtorder 0.15$. This corresponds to ellipticity of $\epsilon\equiv 1 - b/a\gtorder 0.4$, where $2a$ and $2b$ are the bar's major and minor axes. At lower values the bar basically dissolves and represents an oval distortion.

We obtain the bar radius, $R_{\mathrm{bar}}$, by fitting ellipses to the contour map of the face-on disk. The bar radius is determined as the radius where the ellipticity value, $\epsilon(r) = 1 - b/a$, has decreased by 15\% below its maximum. Typically, the position angle of $m=2$ mode stays constant within $R_{\rm bar}$, and changes thereafter --- this can be used to verify $R_{\rm bar}$ \citep[e.g.,][]{laine02}. Our method has been confirmed by a direct calculation of the radial extent of the stellar orbits aligned with the bar and which form its backbone \citep{marti06,villa09}. In comparison, determining the bar radius as the radius of the maximum of $A_2$ \citep[e.g.,][]{algor17} results in neglecting a substantial fraction of the bar stellar and gas masses, as introduces an error in positions of the major resonances.  

Note, that our definition of the bar strength depends on the integrated $m=2$ mode over the bar length (defined below). This differs from alternative method of using the maximum of $A_2(R)$ \citep[e.g.,][]{algor17,rosa20}. The advantage of our method lies in that it depends on the global property of the bar and not on the local one, which can be heavily influenced by the radial mass distribution in the bar.

The deprojection of the galactic disk must be performed carefully in order to maintain the continuity of the bar phase angle as a function of time and to determine the bar pattern speed, $\Omega_{\rm bar}$. We deproject the bar onto three major planes. This is performed by finding the line of nodes of the disk with respect to the plane of the sky, and then rotating the disk with respect to this line of nodes. The pattern speed, $\Omega_{\rm bar}$, is calculated using 5-point time derivative for stability.

The volume of the bar has been determined by using the isodensity contour which determines the bar size, assuming that the bar vertical thickness is the same as that of the stellar disk. The stellar and gas masses of the bars, $M_{\rm bar}$, have been obtained by summing up the gas and stellar particles within this volume.

\section{Results} \label{sec:results}
We start by presenting the evolution of basic parameters which characterize stellar bars in all the models. In order to understand the evolutionary details, one should relate them to the evolution of the parent galaxy itself, and, in some cases, to that of the parent DM halo and associated galaxy interactions, e.g., the cold accretion flow onto the galaxy and the halo, mergers, flybys, halo shapes and substructure. Some of these important details regarding galaxy evolution of our models have been provided in Paper\,I.

Our choice of the end products, i.e., that the DM halos have similar final masses, $M_{\rm vir}$, at the target redshifts of $z_{\rm f}$ = 6, 4, and 2, and the resulting virial radii, $R_{\rm vir}$, in comoving coordinates, have been provided in Table\,\ref{tab:DMsim}. Hence, the average densities of DM halos decrease with decreasing $z_{\rm f}$.

The host galaxies average radii increase with time in physical coordinates: $\sim 1.4$\,kpc at $z_{\rm f}=6$, $\sim 1.6$\,kpc at $z_{\rm f}=4$, and $\sim 3.6$\,kpc at $z_{\rm f}=2$ (Table\,\ref{tab:DMsim}). They span a narrow range of baryonic masses $\sim 3-4\times 10^{10}\,{\rm M_\odot}$ at $z_{\rm f}$. Final stellar masses of galaxies, $M_*$, their radii, $R_{\rm gal}$, and gas fractions, $f_{\rm gas}$, are given in Table\,\ref{tab:DMsim}. Also, Figures\,\ref{fig:starbarmassfrac} and \ref{fig:gasgal} display the evolution of $M_*$ and $f_{\rm gas}$ with time, respectively.

For clarity, we marked the time periods of major mergers in Figure\,\ref{fig:starbarmassfrac}, where we have followed the general notation for major mergers having mass ratio $\gtorder 1:3$, for intermediate mergers $1:4-1:10$, and for minor mergers $\ltorder 1:11$. Note however, intermediate and minor mergers as well as cold accretion can play equally important role in shaping the galactic morphologies.

These galaxy masses mean that the number of stellar particles per galaxy is about $10^6$, and the amount of DM particles per parent halo is about $3\times 10^6$ at $z_{\rm f}$, which is sufficient to resolve dynamical processes within them \citep[e.g.,][]{dubi09}. We limit our analysis to $z\ltorder 9$ for $z_{\rm f} = 6$ objects, to $z\ltorder 7$ for $z_{\rm f} = 4$ objects, and to $z\ltorder 5$ for $z_{\rm f} = 2$ objects, in order to always have a sufficient number of particles per galaxy and per halo. Hence, galaxies which evolved to $z_{\rm f} = 6$ have been analyzed for $\sim 0.9$\,Gyr, those evolved to $z_{\rm f} =4$, for $\sim 1.6$\,Gyr, and galaxies at $z_{\rm f} = 2$, for $\sim 3.3$\,Gyr. 

An important difference between galaxies discussed here and those observed at lower redshifts is the gas fraction. Note that the overall evolution time for our galaxy sample is less than 3\,Gyr, which is a relatively small time period in the life of the $z=0$ galaxies. Our sample galaxies have evolved from the same initial conditions, and differ in their feedback schemes, as this is the main factor which affects their gas fractions. The CW models have about equal fraction of gas and stars for $z_{\rm f} = 6$ runs, and demonstrate a slow but steady decline in $f_{\rm gas}$ to $\sim 20\%-40\%$ for $z_{\rm f}=4$ and 2 models (see Fig.\,\ref{fig:gasgal} and section\,\ref{sec:gasfrac}). On the other hand, a much stronger VW feedback has resulted in suppression of the star formation rate (SFR), and galaxies being dominated by the gas at the level of $\sim 90\%$ initially, with a slow decay over time to $\sim 80\%$.

Galaxy rotation curves are given in Figure\,\ref{fig:galrot}. We observe a gradual decline of the maximal rotational velocity  with time, especially for $z_{\rm f}=2$ galaxies. As we have verified, this is the consequence of the decrease in the central density with time. The galaxy centers display a gradually shallower rotation despite that a shallower central DM cusp is replaced by the baryonic cusp \citep[e.g.,][]{roma08}. 

\subsection{Basic properties of modeled galactic bars}
\label{sec:basic}

The dynamical importance of stellar bars depends in part on their mass fraction with respect to the parent galactic disk. Figure\,\ref{fig:starbarmassfrac} shows that the CW models exhibit massive bars even at the initial redshifts displayed, with their mass fraction starting at $\sim 60\%-80\%$ of the parent stellar galaxy for $z_{\rm f}=6$ objects, and gradually declining to $25\%-30\%$. For $z_{\rm f}=4$ and 2 objects, this mass ratio starts at $30\%-40\%$ and, while increasing or decreasing with time, ends basically unchanged at $25\%-30\%$. Occasional sudden increases or decreases of this fraction have been checked by us and have been found always related to the environmental effects: mergers, close flybys and strong gas inflows along the extragalactic filaments. 

The VW models have lower mass ratios, $\sim 30\%-70\%$ for $z_{\rm f}=6$ bars, initially, exhibiting substantial variability, either declining or increasing to 50\%. $z_{\rm f}=4$ and 2 bars  display variability and either decline, increase or stay unchanged with time. This variability is closely related to various types of galaxy interactions, as we discuss in detail in the following subsections. At $z_{\rm f}$, the bar mass fraction, $\sim 25\%-60\%$, appears to be higher than in the CW models.

\subsubsection{Evolution of bar amplitudes}
\label{sec:amplitude}

\begin{figure}
\center 
	\includegraphics[width=0.7\textwidth]{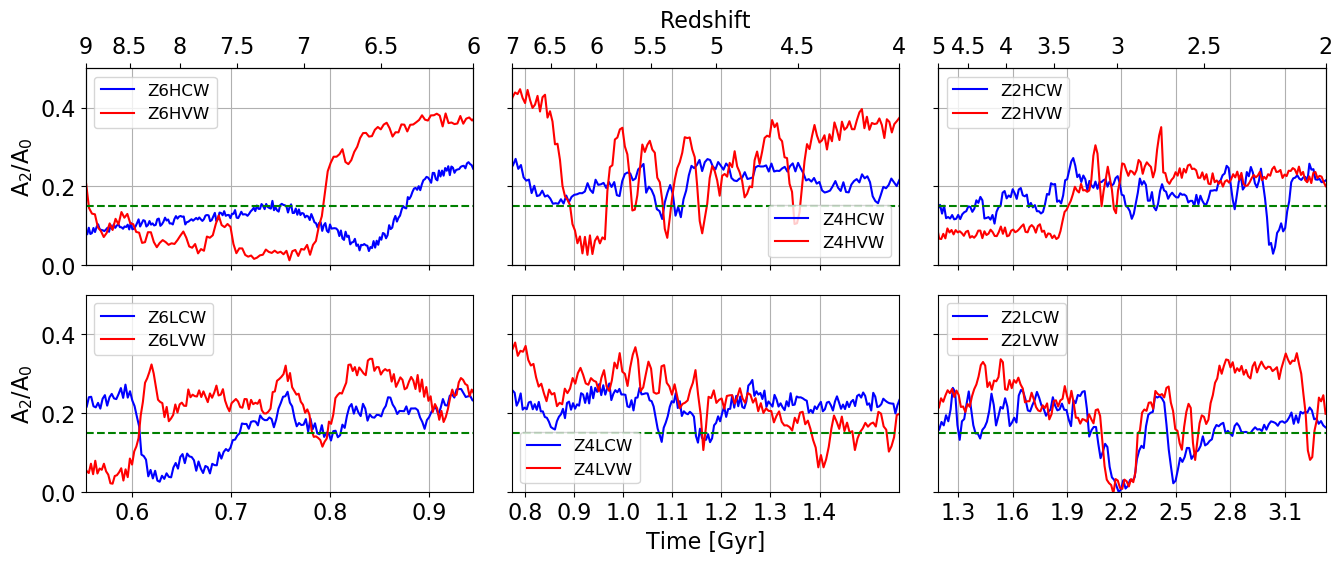}
    \caption{Evolution of normalized Fourier amplitude, $A_2$, of stellar bars by the monopole term, $A_0$, for the sample galaxies describing the evolution of the bar strength. The VW models are given by the red lines, and CW models by the blue lines. The green dashed lines represent $A_2/A_0=0.15$, which is the border line between bars and oval distortions.}
    \label{fig:a2}
    \end{figure}

The stellar bar $m=2$ mode Fourier amplitude evolution for our models are shown in Figure\,\ref{fig:a2}. For isolated galaxies the values of $A_2$ obtained in numerical simulations typically reach $\sim 0.5$ \citep[e.g.,][]{villa09,dubi09}. $A_2\ltorder 0.15$ are typically referred to as oval distortions. In the following we adopt these definitions and consider this value as a threshold for $m=2$ mode which we call a stellar bar. Hence, we consider the bar forming when the amplitude of the $m=2$ model becomes larger than 0.15, and it is destroyed when its amplitude dives below this threshold.

In the following, we describe the main events which lead to the major changes in the bar amplitudes in Figure\,\ref{fig:a2}, but due to the large number of these events we refrain from displaying their analysis. Instead, we present as examples 3 different events in section\,\ref{sec:omegabar} and illustrate the associated processes which trigger them in detail.

Gas-rich stellar bars have developed in all our models. Rarely they appear strong. The bar sizes typically increase towards lower redshifts, in tandem with the disk growth. The longest bars appear after $z\sim 2.5$. The VW models host longer bars towards $z_{\rm f}$, with the exception of $z_{\rm f}=2$ galaxies, where at the end of the simulations the CW and VW bars are essentially having the same sizes.

The Z6HCW bar has formed only at $z\sim 6.3$. As a result of a merger, the amplitude of $m=2$ mode increases substantially to $\sim 0.3$. The associated VW model for the same halo has no bar until it experiences multiple mergers around $z\sim 6.9-6.5$, which trigger a strong bar with a jump in the amplitude to $\sim 0.3$. Interestingly, some of these mergers are retrograde and some prograde and the effects of their actions are expected to be opposite. The resulting bar reaches the amplitude of $\sim 0.4$ towards $z\sim 6$. This is a very strong bar. 

The bar in the model Z6LVW, behaves differently. This bar is triggered by a prograde, low-inclination merger at $z\sim 8.5$ and becomes strong, $A_2\sim 0.35$, but the galaxy experiences a retrograde merger at $z\sim 8.2$ which weakens it to $A_2\sim 0.2$. Its strength remains the same until $z\sim 7.3$, when it is strengthened by a substantial gas inflow along a prograde gas filament. A retrograde encounter at $z\sim 7$ weakens it to an oval distortion, destroying the bar. Subsequent prograde encounter reforms the bar with $A_2\sim 0.32$. It decays gradually and strengthens again, all within the range of $A_2\sim 0.2-0.3$.

A weak bar develops already before $z\sim 9$ in the associated Z6LCW model, appearing as a response to the overall asymmetry in the galactic disk. This asymmetry originates when the galaxy mass is slightly below $\sim 10^8\,{\rm M_\odot}$ apparently because of the penetrating close encounter which distorts the parent DM halo and coincides with a strong gas inflow. The bar is destroyed abruptly at $z\sim 8.4$. Thereafter, the galaxy experiences sporadic gas inflows along the penetrating filaments, and the bar reappears around $z\sim 7.5$, with a strength oscillating around $\sim 0.2$.  

Galaxies with $z_{\rm f} = 4$ continue this trend with the bar amplitude oscillating in the range of $\sim 0.12-0.35$ ($z\sim 6-4.4$), in Z4HCW. Thus the bar comes close to being destroyed, but recovers. Its counterpart in Z4HVW displayed even stronger amplitude variations, undergoes multiple destructions and reformations. Exceptionally strong bar in this model actually exists at higher redshifts, $z\sim 6.5-7$. 

Bars in the Z4L halo behave somewhat differently. The CW model hosts a very weak bar which is destroyed around $z\sim 5.2$, and reforms shortly after this. In VW model, a relatively strong bar forms at redshifts $> 7$, and slowly decays, being destroyed around $z\sim 5.1$. These amplitude oscillations are correlated with the ongoing intermediate mergers, some prograde and some retrograde. 

\begin{figure}
\center 
	\includegraphics[width=0.7\textwidth]{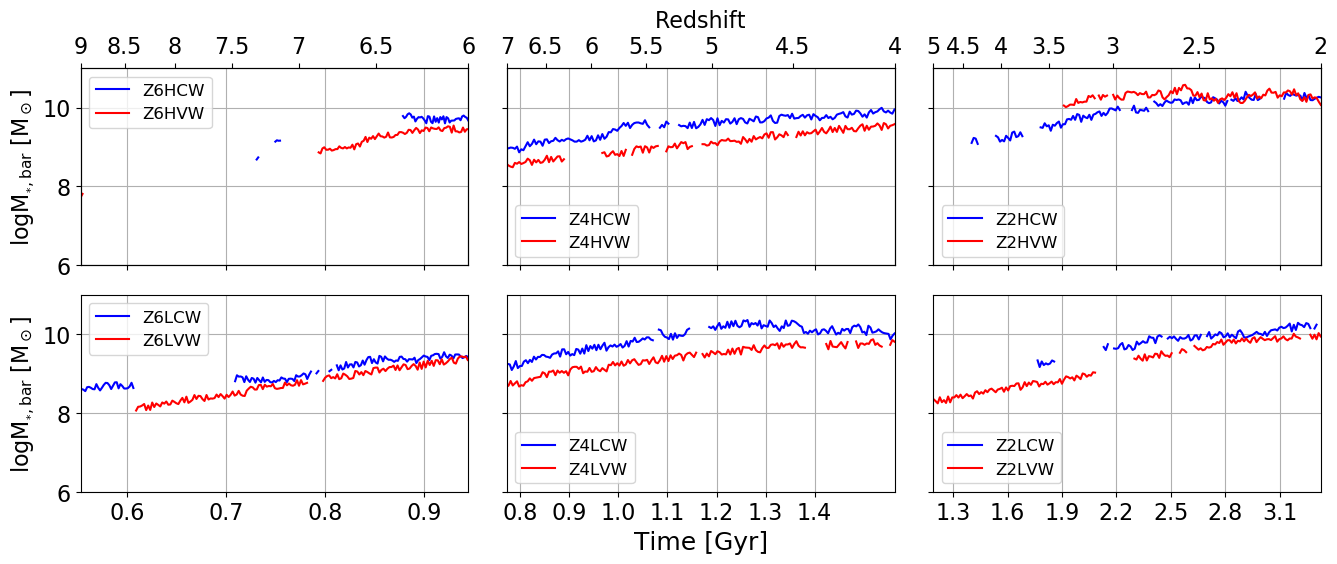}
    \caption{Evolution of stellar bar masses for modeled galaxies. The VW models are given by the red lines, and CW models by the blue lines. Only time periods when the bars exist, i.e., $A_2/A_0\gtorder 0.15$ in Figure\,\ref{fig:a2}, are shown here. The gaps correspond to the bar dissolution times.}
    \label{fig:barmass}
    \end{figure}

Galaxies with $z_{\rm f} = 2$  display a more stable evolution, with $m=2$ amplitudes varying in the range of $0.1-0.35$, occasionally dropping below 0.15, so being destroyed. We note that mergers and close encounters perturb the bar positions with respect to the disk center in all models, resulting in a bar which oscillates around the common center of mass. Oval distortions are present in all models. We only analyze the time periods when the stellar bars are present and display their stellar mass evolution in Figure\,\ref{fig:barmass}.

To summarize the bar amplitude evolution in these models, we note that stellar bars go through multiple formations and destructions events. This behavior is very different from evolution at low $z$, in isolated models and galaxies outside the clusters of galaxies environment. The VW bars are typically stronger than the CW bars, but opposite case are not rare as well. Note, as we show in Figure\,\ref{fig:gasbar}, the VW bars are more gas-rich and even gas-dominated. The gas reacts stronger to any non-axisymmetric perturbation, and "drags" the stellar component along. 

\begin{figure}
\center 
	\includegraphics[width=0.4\textwidth]{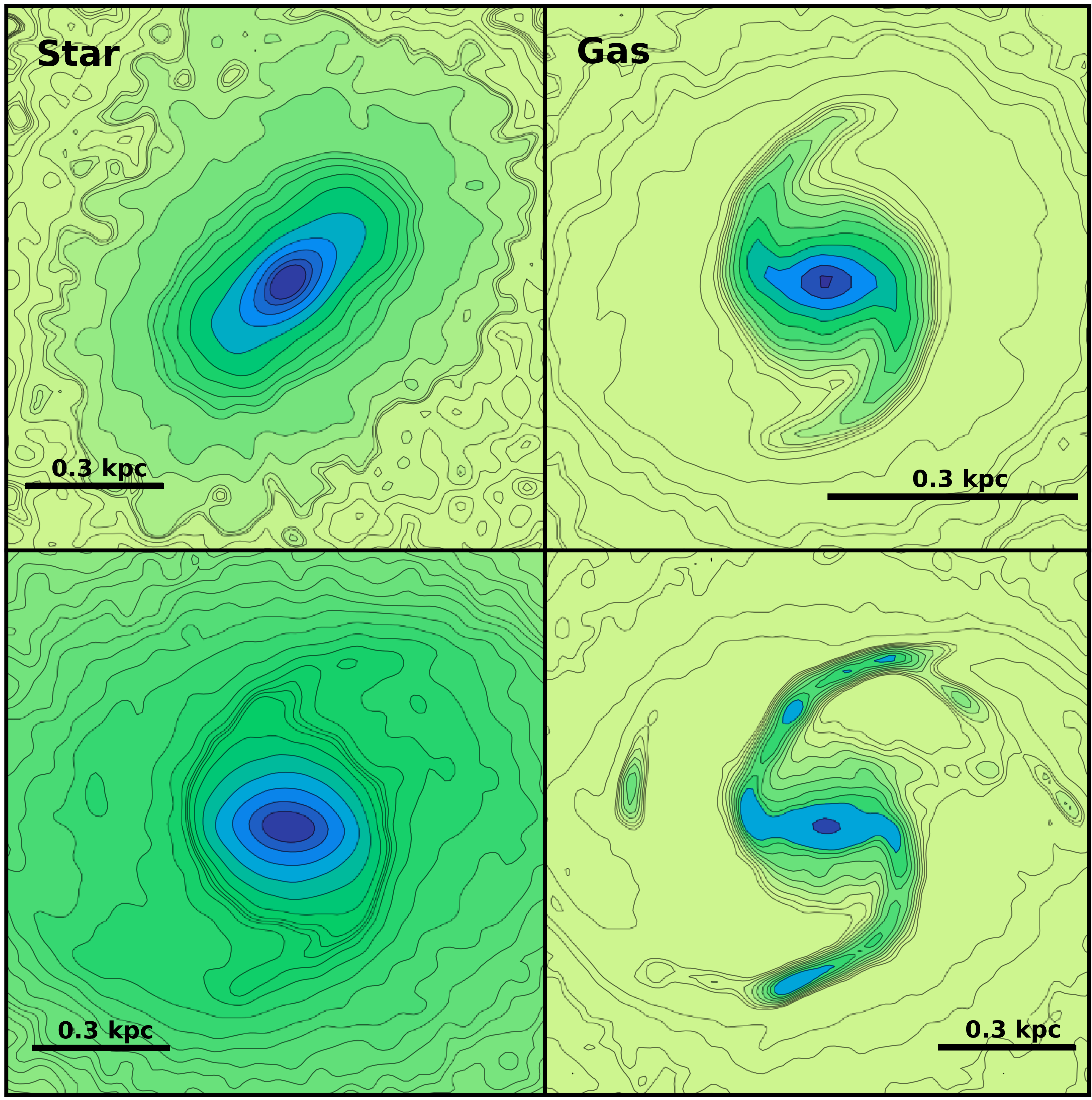}
    \caption{Examples of face-on stellar bars and their gaseous components. Contours represent the stellar or gaseous surface density, and are logarithmically spaced. Stellar bars are shown on the left and the gaseous ones are on the right. Left (top to bottom): Z2HVW at $z=4.35$ with characteristic boxy shapes of outer contours extending to the UHR (see text); Z4HCW at $z=4.36$ with characteristic oval shapes of contours. Right: (top to bottom): Z4HCW at $z=6.7$; Z4HCW at $z=4.46$.}
    \label{fig:shapes}
    \end{figure} 
 
 \begin{figure}
\center 
   \includegraphics[width=0.4\textwidth]{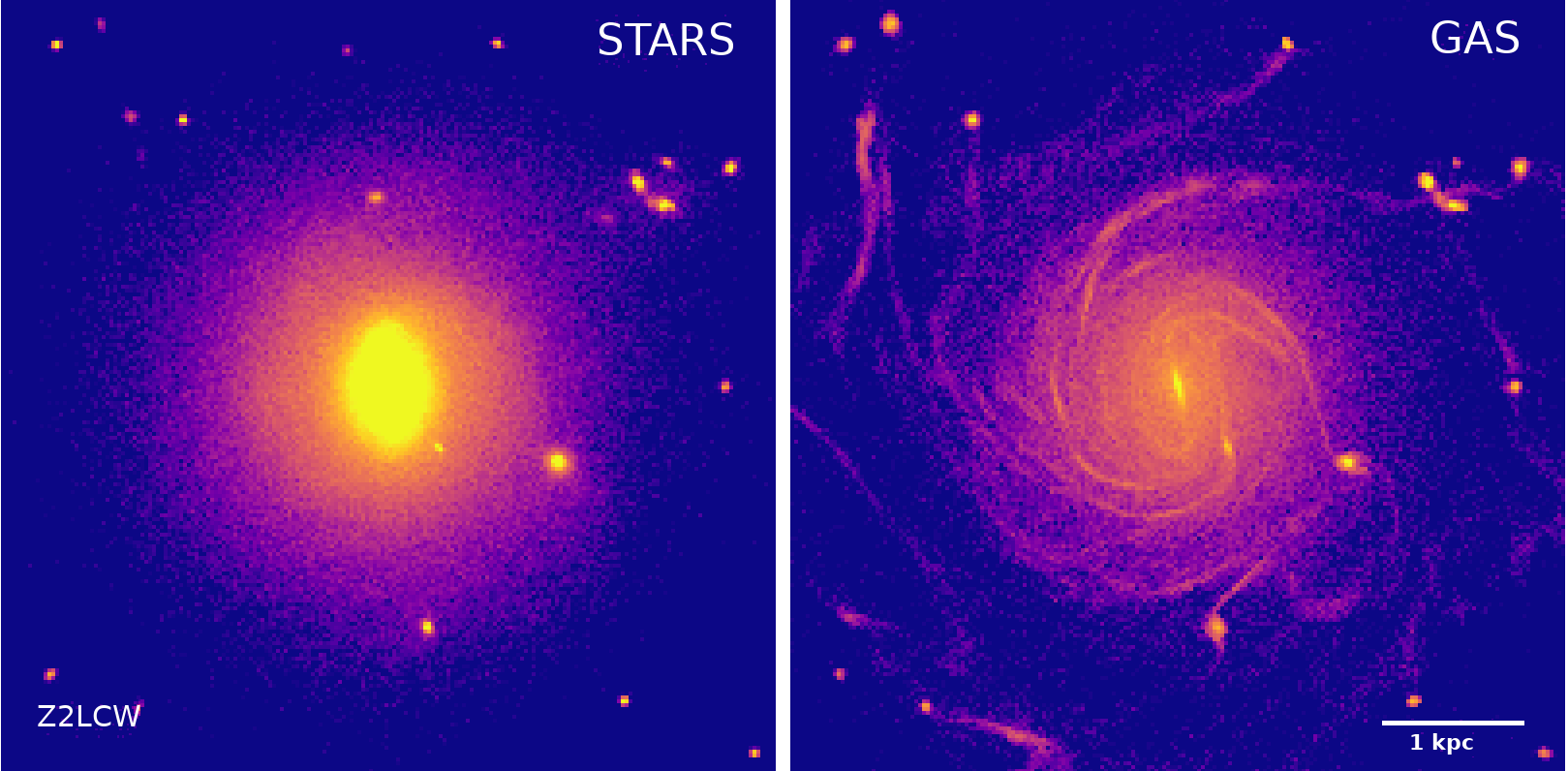}
    \caption{Face-on galaxy Z2LCW at $z=2$ with stars (left frame) and gas (right frame) colored with projected surface mass density. Note the central prominent bar and the two spiral arms it drives. The stellar disk is relatively "hot" (large dispersion velocities) and hence the stellar spirals are barely visible, but appear prominent in the gas. }
    \label{fig:gal_stars_gas}
    \end{figure} 
 
\subsubsection{Bar shapes} 
\label{sec:shapes} 
 
The bar shapes depend on the outermost trapped orbits by the bar. Strong bars extend to the Ultra-Harmonic resonances (UHR), located just inside the CR, and have characteristic boxy shapes associated with the outer 4:1 orbits, i.e., four radial oscillations during one rotation. An example of such bars is shown in Figure\,\ref{fig:shapes}. We have displayed two basic face-on bar shapes observed in our numerical simulations, i.e., boxy- and ovally-shaped when viewed face-on (Z2HVW and Z4HCW, respectively). We also did find bars with characteristic boxy/peanut shaped bulges when viewed edge-on, discuss their low-$z$ counterparts in section\,\ref{sec:locals}, and defer the explanation to section\,\ref{sec:discussion}.  The face-on galaxy Z2LCW with a typical bar in stellar and gaseous components is shown in Figure\,\ref{fig:gal_stars_gas}.
 
We have observed also some cases of the gaseous component in the bar being elongated nearly perpendicular to the stellar component in the bar, but do not show it here. This phenomenon is associated with the appearance of the Inner Lindblad resonance (ILR) within the bar \citep[e.g.,][]{sell93,shlo99}. Detection of this response in the gaseous component to the driving by the bar indicates that we have resolved the dynamics within the bar accurately enough.

\subsubsection{Evolution of bar pattern speeds} 
\label{sec:omegabar} 
 \begin{figure}
\center
\includegraphics[width=0.7\textwidth]{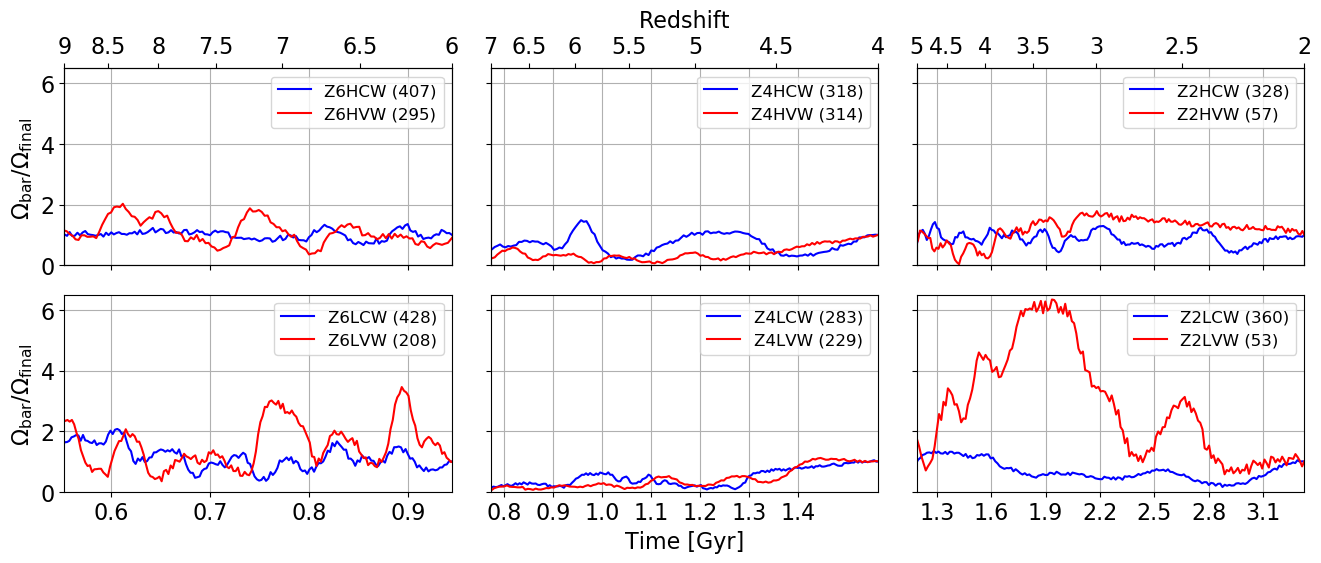}
    \caption{Evolution of stellar bar pattern speeds, $\Omega_{\rm bar}$, normalized by the final redshift pattern speeds $\Omega_{\rm final}(z_{\rm f})$, at $z_{\rm f} = 6, 4$, and 2. The VW models are given by the red lines, and the CW models by the blue lines.  $\Omega_{\rm final}$ has the values in parentheses (all in units of ${\rm km\,s^{-1}\,kpc^{-1}}$).
    }
    \label{fig:omega}
    \end{figure}

 \begin{figure}
\center
\includegraphics[width=0.7\textwidth]{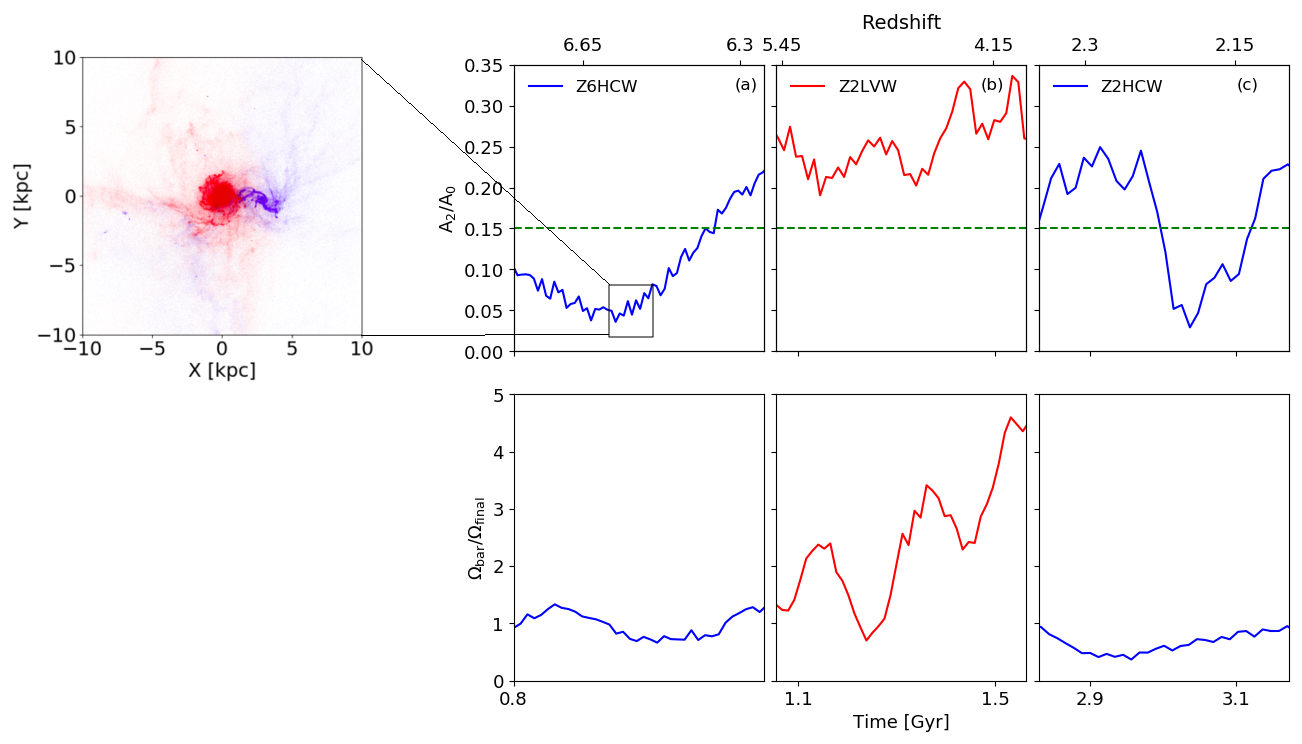}
    \caption{Three examples aimed at specific galaxy interactions: shown is evolution of the bar amplitude (top frames) and the bar pattern speed (lower frames). The dashed line at $A_2/A_0=0.15$ separates between the bar and the oval distortion. (a) Model Z6HCW during asymmetric gas inflow to the galaxy. This interaction triggers the bar and gradually speeds it up. The insert displays the prograde (red, the galaxy) and retrograde (blue, the inflow) velocities of the gas at $z\sim 6.55$. (b) Model Z2LVW series of 3 prograde and nearly complanar minor mergers. These interaction increase the bar strength and its pattern speed. (c) Model Z2HCW retrograde major merger which destroys the bar, which is later triggered anew.}
    \label{fig:example}
    \end{figure}

We have normalized the bar pattern speeds, $\Omega_{\rm bar}$, by their final values at $z_{\rm f}$. This was done for an easier comparison of $\Omega_{\rm bar}$ evolution based on our model construction --- so at $z_{\rm f}$, the normalized $\Omega_{\rm bar}$ always tends to unity (Fig.\,\ref{fig:omega}).  

The most striking feature of $\Omega_{\rm bar}$ in our models appears to be its variability. At low redshifts, and especially in isolated models, the bar pattern speed is a monotonically decreasing function of time. This is understandable, as stellar bars experience friction and channel their angular momentum to the outer disks, outside the corotation, and mostly to the parent DM halos. This behavior is not noticeable at high redshifts. In our models, the bar pattern speed evolution is governed by galaxy interactions with its environment. Prograde and retrograde mergers and flybys, as well as the cold accretion along the filaments, completely dominate the angular momentum redistribution and hence the pattern speed amplitude variation of stellar bars. Such oscillations of  $\Omega_{\rm bar}$, with prolonged periods of increase, have been never observed at low redshifts. In section\,\ref{sec:barpop}, we correlate this behavior with the star formation processes in the galaxy and in the bar.

 \begin{figure}
\center 
	\includegraphics[width=0.7\textwidth]{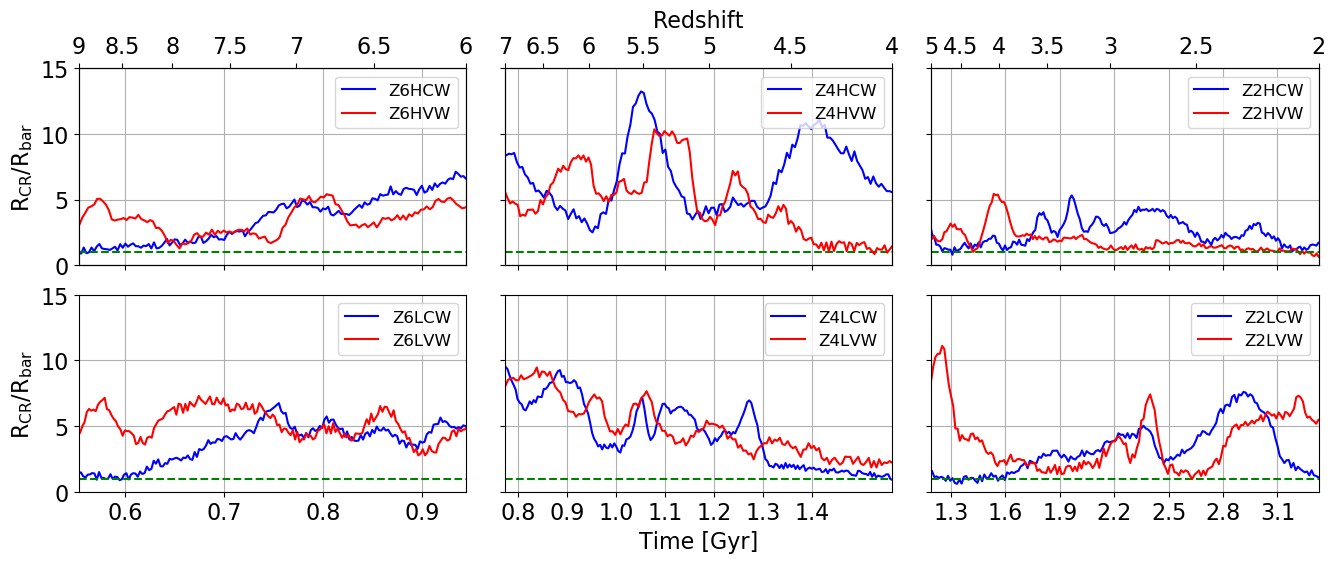}
    \caption{Evolution of the CR-to-bar radius ratio, $R_{\rm CR}/R_{\rm bar}$, of stellar bars. The VW models are given by the red lines, and CW models by the blue lines.  The green dashed lines represent $R_{\rm CR}/R_{\rm bar}=1$.  }
    \label{fig:CRtoRb}
    \end{figure}

 \begin{figure}
\center 
	\includegraphics[width=0.7\textwidth]{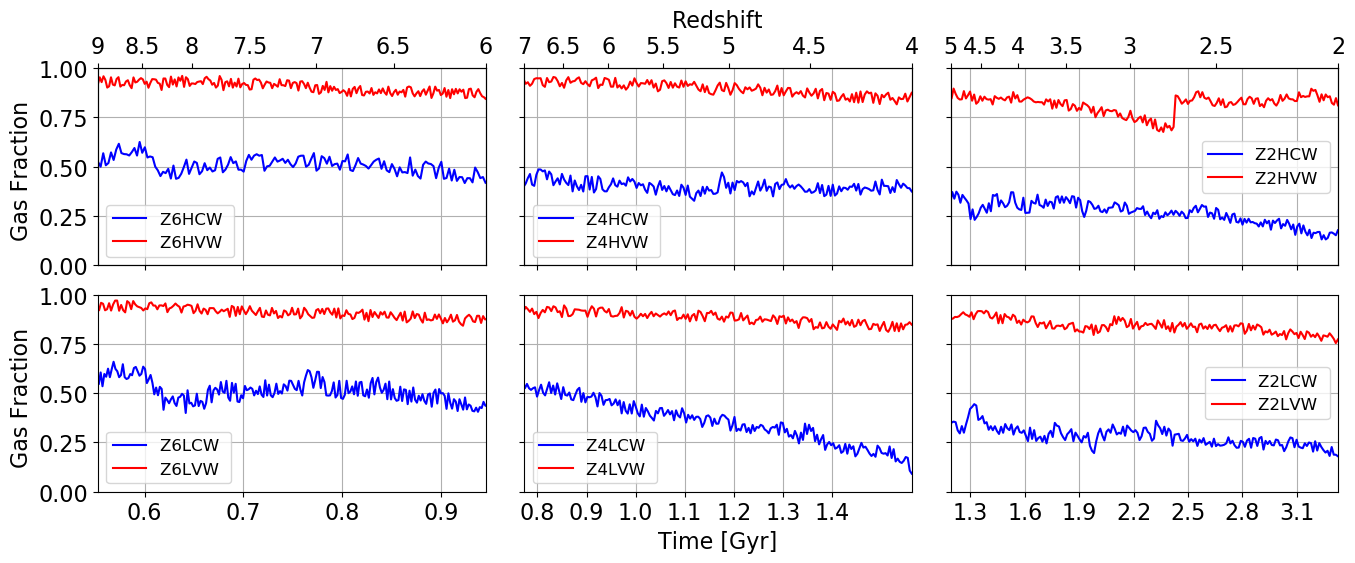}
    \caption{Evolution of gas fraction in the host galaxies. The VW models are given by the red lines, and CW models by the blue lines.}
    \label{fig:gasgal}
    \end{figure}

Overall, the bar pattern speeds appear not to depend on redshift, except their superposed variability. This variability is very individual to the galaxy models which lie in similar halos and environments, e.g., variability in the pattern speeds of bars in Z6HCW and Z6HVW. While averaging $\Omega_{\rm bar}$ over time in these two galaxies results in very similar averages, $<\Omega_{\rm bar}>$, the average pattern speeds for the Z6LCW and Z6LVW pair differ by about 50\%. Other pair of CW and VW models exhibit average $<\Omega_{\rm bar}>$ which differ not more than $\sim 30\%$. Extreme variability can be observed in the pattern speed of Z2LVW galaxy.

Bars predominantly brake against the DM in isolated models, but in the models presented here the gas inflow rates can fuel a different response from the bars. For example, in the gas-rich models (i.e., VW galaxies), the gas appears to be channeled to the center which results in balancing the braking or even accelerating the bars altogether, in a complete agreement with the gas-rich models analyzed by \citet{villa10}. 

At the same time, for each model, the bar pattern speeds were found to correlate with galaxy interactions. In this list, one can include the last stages of mergers, close flybys, and cold filamentary and diffuse accretion. Prograde mergers with a low inclination to the disk plane always lead to the speed-up of stellar bars in our models, but we do not find the same correlation with their amplitudes. A similar behavior has been observed for close low-inclination prograde flybys. Retrograde mergers and flybys slow down the bars, but their amplitudes show both increase and decrease, frequently below the threshold of $A_2\sim 0.15$. 

This behavior differs from that of the isolated models (typically without the gas component). In these models a clear (anti)-correlation is observed, i.e., the bar slowdown is always associated with increase in $A_2$, and vice versa. The reason for this is that the bar slowdown is the result of the angular momentum transfer mostly to the DM, when both the tumbling angular momentum and the internal circulation are reduced, leading to a more radial orbits and increased bar ellipticity. Speed up of the bar comes from increased angular momentum both in tumbling and in the internal circulation, this leads to a decreased ellipticity of the bar.

We also observe the bar slowdown when mergers and gas influx have high inclinations to the disk plane. Finally, the filamentary accretion which comes in the form of both diffuse and clumpy components, can join the galactic disk in a prograde or retrograde fashion. Mergers and gas influx change the galaxy mass concentration and therefore its rotation curve. All this has a direct impact on the kinematics of stellar bars at these redshifts.

We give three detailed examples of the bar amplitude and its pattern speed response to the asymmetric gas inflow, and prograde and retrograde mergers in Figure\,\ref{fig:example}.

Most prominent example of multiple gas-rich minor mergers and gas influx can be observed in $z_{\rm f}=6$ models. In Z6HCW model (around $z\sim 6.65-6.3$, see Fig.\,\ref{fig:example}a), we detect a substantial asymmetric gas influx into the disk. Prior to this event, we measured only a weak oval distortion in the stellar disk with $A_2\sim 0.05$, so the disk was nearly axisymmetric. This gas influx triggered the stellar bar formation. The oval's pattern speed that was decaying prior to this event experienced an increase thereafter. A similar event happened in Z6HVW at $z\sim 6.9$.

Two other models, Z6LCW display a single intermediate gas-rich merger followed by a number of minor prograde and retrograde mergers, after  $z\sim 7.5$. While in Z6LVW we observe a large gas influx followed by multiple minor mergers at $z\sim 8.4$. The bar formation has been triggered by these events, and always associated with an increase in the pattern speed, as each event is spread out in time over a few bar rotations.

\begin{figure}
\center 
	\includegraphics[width=0.7\textwidth]{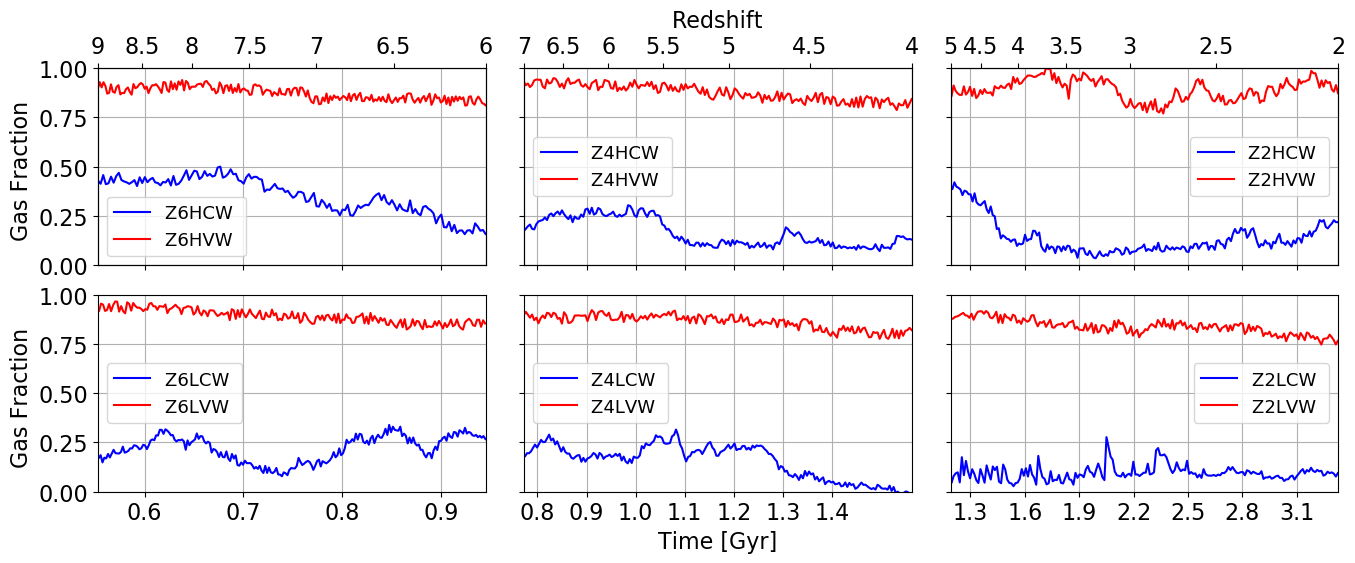}
    \caption{Evolution of the gas fraction within stellar bars.  The VW models are given by the red lines, and CW models by the blue lines.}
    \label{fig:gasbar}
    \end{figure}

Models with $z_{\rm f}=4$, exhibit stellar bars for more prolonged time periods than $z_{\rm f}=6$ models. For example, Z4HCW hosts the bar almost all the time, except around $z\sim 5.5-5.3$, when retrograde minor merger terminates it (i.e., drives its amplitude below the threshold), as well as after $z\sim 4.7$ (influx of retrograde minor mergers and gas). Z4HVW displays the bar after $z\sim 5.9$, when a dramatic gas influx is followed by prograde and retrograde minor mergers, leading to large amplitude oscillations in $\Omega_{\rm bar}$ and especially in $A_2$. The Z4LVW galaxy displays the bar until $z\sim 4.4$, when a dramatic event of a nearly vertical merger puffs up the stellar bar and the disk as well as its gaseous component. Note, as shown by Figure\,\ref{fig:gasbar}, the gas component dominates over the stellar bar component all the time at these $z$. The heating of gas in the bar is contributed by the burst of star formation associated with this event (Fig.\,\ref{fig:sfrbar}). Therefore, heating by the vertical merger and spreading out of the gas has led to thickening of the stellar components and reduction of stellar bar amplitude below the threshold.

Next, models with $z_{\rm f}=2$ have the longest sustainable stellar bars, albeit weak ones. The most interesting is the evolution of Z2LVW, which experienced a series of nearly coplanar minor mergers which bust the stellar bar strength between $z\sim 5.45-4.15$, and dramatically increase its pattern speed by a factor of $\sim 3$, as shown in Figure\,\ref{fig:example}b. Moreover, at $z\sim 3.1$, a high-inclination merger damps the bar completely, while after $z\sim 2.9$ the bar has been triggered again. Around $z\sim 2.1$, two retrograde minor mergers weaken the bar for a short time period.

Finally, Z2HCW hosts the bar intermittently over of the displayed time period. After $z\sim 2.3$ it experiences a retrograde merger which comes in at
low inclination orbit. The stellar bars is abruptly destroyed, with $A_2$ falling below 0.05. It is triggered again around $z\sim 2.15$. During the retrograde merger, the bar pattern speed slows down, then starts to grow, as shown in Figure\,\ref{fig:example}c. Z2LCW exhibits the bar at $z\sim 3.6-3.4$ and after $z\sim 3.1$. This bar has been terminated at the very end by a merger.  

To summarize, a high variability in basic properties of stellar bars, especially in their amplitude and pattern speed, is characteristic of high-redshift evolution and is dominated by the galactic environment. The effect of these interactions have been modeled and analyzed in low-redshift and isolated models \citep[e.g.,][]{ger90,bere04,roma08}. What appears to be novel at high redshifts is the high frequency and even overlapping of these events which completely overwhelms the traditionally monotonic decrease in the bar pattern speed. 
 
\subsubsection{Evolution of the CR-to-bar radius ratio, $R_{\rm CR}/R_{\rm bar}$}
\label{sec:CRtoRbar}

Figure\,\ref{fig:CRtoRb} displays the evolution of the $R_{\rm CR}/R_{\rm bar}$ ratio in our models. Models targeting $z_{\rm f}=6$ show this ratio is generally increasing with time from the 1--4 range to about 5--7.  $z_{\rm f}=4$ models  exhibit a different behavior, either $R_{\rm CR}/R_{\rm bar}$ decays with time from $\sim 8$ to $\sim 1$, or broadly oscillating between 3 and 13 (for CW) and declining to 1.5 at the end of the simulations (e.g., in the VW models). Lastly, $z_{\rm f}=2$ models show this ratio oscillating in the range of 1--5 and 1--10 (CW), or oscillating between 1  and 10 (VW).
In all cases, $R_{\rm CR}/R_{\rm bar}\gtorder 1$ at all times. 

\begin{figure}
\center 
	\includegraphics[width=0.7\textwidth]{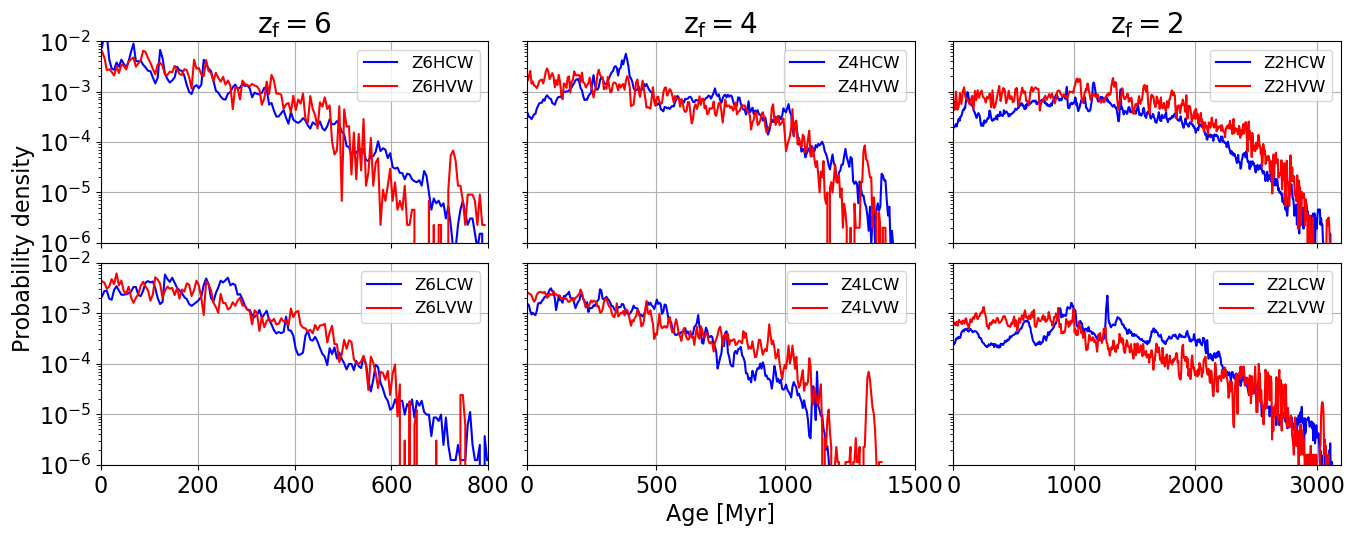}
    \caption{The PDF of stellar ages in bars.  The time shown is the stellar age for each model  with respect to $z_{\rm f}$. The curves have been normalized by the total number of stars in the parent galaxies at $z_{\rm f}$. The ages were divided into bins of 5\,Myr. The VW models are given by the red lines, and CW models by the blue lines.}
    \label{fig:ages}
    \end{figure} 

As the galaxies are always small at these redshifts, the stellar bars have sub-kpc sizes (in physical coordinates). The average $R_{\rm CR}/R_{\rm bar}$ in our models declines with $z_{\rm f}$, from $\sim 5.35$ at $z_{\rm f}=6$ to $\sim 2.32$ at $z_{\rm f}=4$ and to $\sim 2.12$ at $z_{\rm f}=2$. Typically, this ratio for the CW bars is larger than for the VW bars. The reason for the large amplitude oscillations of $R_{\rm CR}/R_{\rm bar}$ is related to variations in $R_{\rm CR}$ and $R_{\rm bar}$ which are not correlated among themselves. We have traced this variability in $R_{\rm CR}$ and $R_{\rm bar}$, separately. The CR radius depends on the bar pattern speed and on the rotational velocity of the parent disk. Both can vary strongly during all types of mergers, flybys and gas redistribution in the disk. The bar size depends on the efficiency of angular momentum exchange between the disk-bar and the DM halo. One expects little correlation between these processes. We return to $R_{\rm CR}/R_{\rm bar}$ variability in section\,\ref{sec:discussion}.   

\subsubsection{Evolution of the gas fraction in bars and host galaxies}
\label{sec:gasfrac}

Figures\,\ref{fig:gasgal} and \ref{fig:gasbar} display the gas fraction, $f_{\rm gas}$, evolution in the host galaxies and their stellar bars, respectively. The methods to determine the bar sizes and masses and the host galaxy parameters have been outlined in sections\,\ref{sec:sims} and \ref{sec:bars}. We find up to about 90\% gas fraction in VW galaxies and up to 60\% in the CW galaxies, initially. These fractions decline mostly slowly and monotonically in VW galaxies towards 75\%--80\%, and sometimes faster in the CW galaxies, down to 20\%--25\%. 

The gas fraction in VW bars appears equally high to their host galaxies during the entire evolution time, but can vary much more in the CW bars. By the initial redshifts shown in Figures\,\ref{fig:gasgal} and \ref{fig:gasbar}, both galaxies and their bars in CW models display gas fractions which are lower by a factor of two compared to the VW runs. Subsequent evolution of CW bars can either show flat but variable $f_{\rm gas}$ or decline with time. The difference between the CW and VW galaxies and bars is consistent with the feedback strength in CW and VW models (section\,\ref{sec:winds}) and the associated star formation rates (SFRs), see section\,\ref{sec:barpop}. 

The decline in gas fraction in  CW bars appears to be somewhat more prominent in the high $\updelta$ models, especially in Z6HCW and Z2HCW. In the latter case, $f_{\rm gas}$ declines by a factor of 4 around $z\sim 4.4$, and is related to the action of an intermediate merger which proceeds strictly in the galactic plane. The satellite spirals in slowly and affects both the gas in the bar (which is pushed out) and the bar size (which decreases substantially as the satellite moves in).

On Z4LCW bar, the gas fraction declines to very low values. On the other hand, in Z6LCW bar, the gas fraction oscillates largely around the initial value of $\sim 25\%$. 

Overall, such gas-rich bars as in VW models are unknown in the contemporary universe, while CW bars are known observationally to be in excess of 70\% in disk galaxies. Below, we shall compare the gas evolution in the bars with other evolutionary parameters.

\subsubsection{Evolution of stellar population in bars}
\label{sec:barpop}

The stellar age distribution within the bar is shown in Figure\,\ref{fig:ages}, and is very similar to the stellar age distribution in the parent galaxies (Fig.\,16 in Paper\,I). The resulting PDFs have been normalized by the total number of stars in galaxies at $z_{\rm f}$. The evolutionary curves display a bursting behavior, which dominates in all models. The PDF curves decay with increasing age. The decay time increases with decreasing $z_{\rm f}$, from $\sim 0.4$\,Gyr in $z_{\rm f}=6$ runs, to $\sim 0.8$\,Gyr for $z_{\rm f}=4$ models, to $\sim 2.2$\,Gyr for $z_{\rm f}=2$ models. The lower PDF value at the age = 0 corresponds to a lower SFR.

Three types of behavior can be observed in the distribution of stellar ages in Figure\,\ref{fig:ages}. First, when most of the stars have been formed close to $z_{\rm f}$ (e.g., top left frame in this Figure). Second, when a plateau has been reached within $0.2-1$\,Gyr before the galaxies have reached the corresponding $z_{\rm f}$ (e., right two frams in this Figure, with VW models). And third, when the peak in stars of a specific age has been reached within $\sim 1$\,Gyr of $z_{\rm f}$, and numbers of younger stars declined thereafter (e.g., top middle and right frames, for CW models). 

The low probabilities shown on the y-axes are due to the high temporal resolution used, i.e., the binning of 5\,Myr. This allows us to estimate a typical burst duration of about 100\,Myr, although some bursts are as long as 100-500\,Myr. Those have been triggered by interactions and flybys and their duration is closely related to the timescale and parameters of encounters.

The distribution of stellar ages in bars should be compared with the evolution of stellar bar-to-galaxy stellar mass ratios in Figure\,\ref{fig:starbarmassfrac}. For example, this ratio is $\sim 0.37$ for Z6HCW at $z_{\rm f}=6$ (Fig.\,\ref{fig:starbarmassfrac}), and it is equal to the integral of the curve in Figure\,\ref{fig:ages} from the initial time of the run till the end of the run. 

\begin{figure}
\center 
	\includegraphics[width=0.7\textwidth]{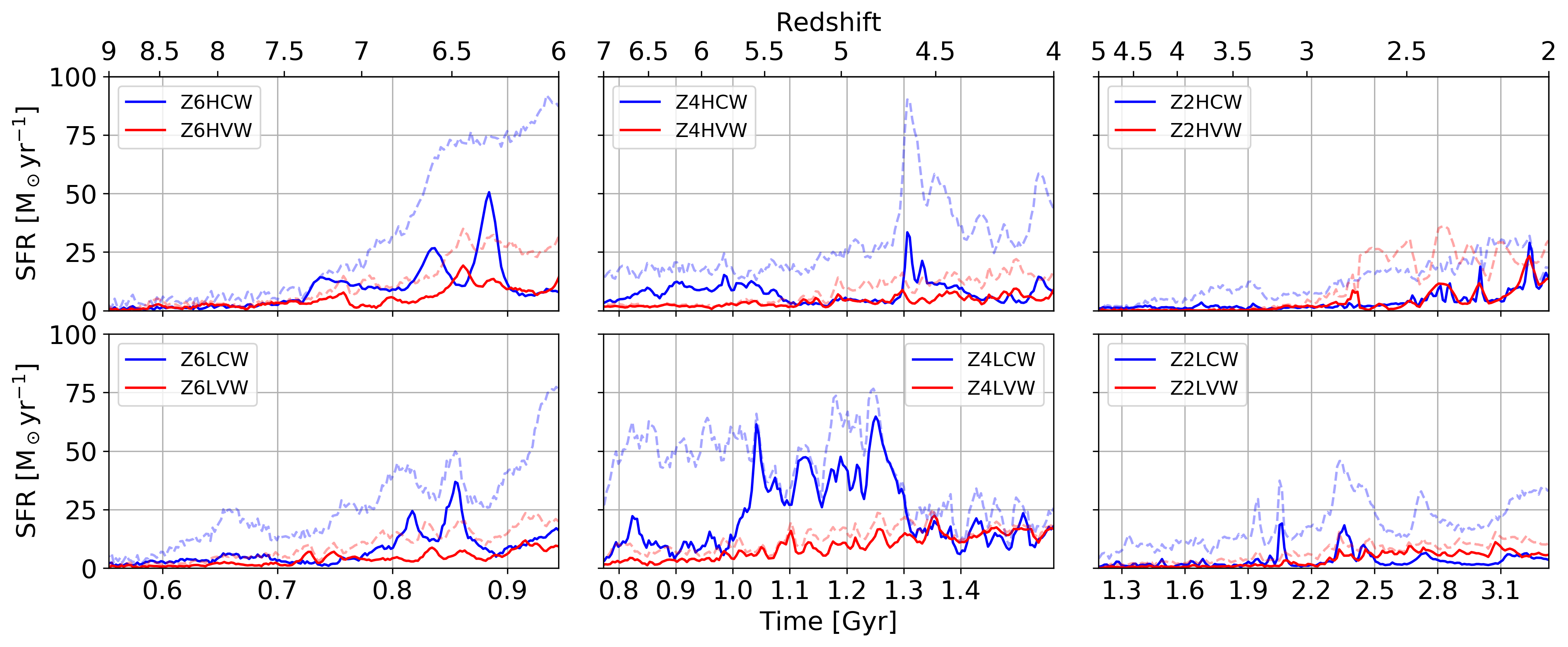}
    \caption{Evolution of star formation rate for stellar bars (thick solid lines) and galaxies (weak dashed lines). The SFR has been calculated based on time intervals of 1\,Myr. The VW models are given by the red lines, and CW models by the blue lines. }
    \label{fig:sfrbar}
    \end{figure}

Note, that there is no one-to-one correspondence between the stellar age distribution in bars and the SFR rate in bars shown in Figure\,\ref{fig:sfrbar}. The reason for this is that ages represent the stars which are still inside the bars at $z_{\rm f}$, while the SFR is given for stars that populated the bar at the time of their formation. Stars could move out or be captured by the bar, and the bar can change its shape and size, thus leaving some stars outside or bringing them in.

While majority of the stars in our bars are very young at the end of the run for $z_{\rm f}=6$ models, and even in some $z_{\rm f}=4$ models, other models, e.g., Z4LVW, have an older population. Moreover, models with $z_{\rm f}=2$ have populations peaking around 1\,Gyr, still a younger population is present in substantial numbers. So, one can state that stellar populations in these gas-rich bars age with decreasing $z_{\rm f}$.

 \begin{figure}
\center 
	\includegraphics[width=0.7\textwidth]{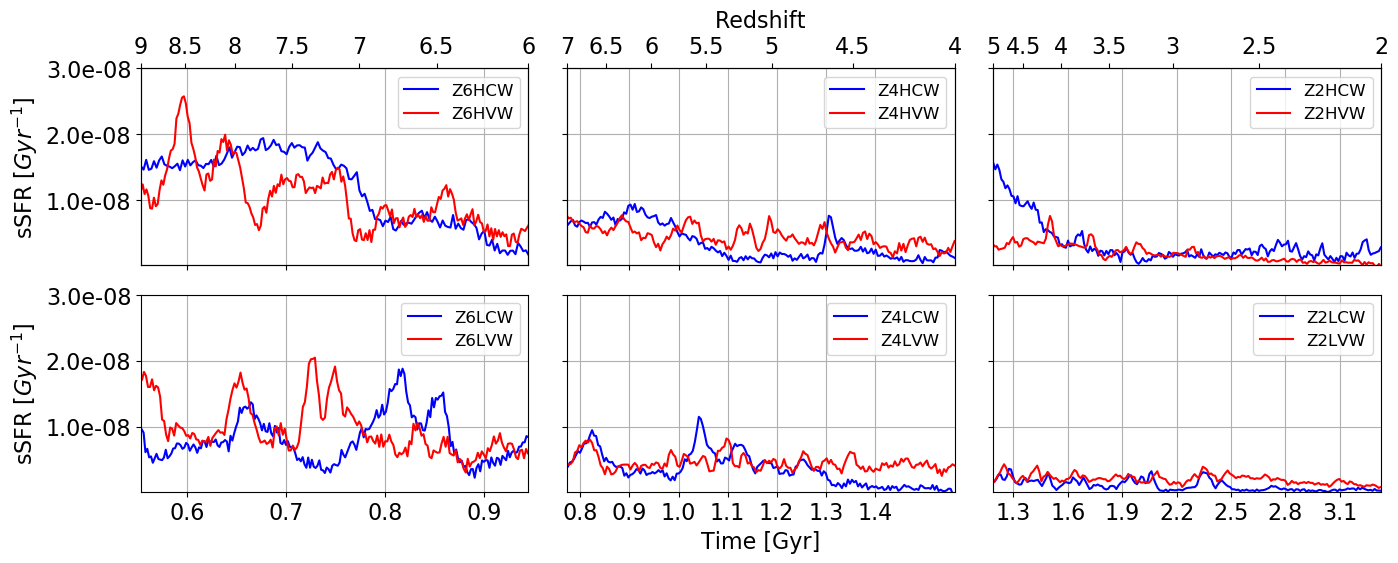}
    \caption{Evolution of specific star formation rate, sSFR, within bars.  The VW models are given by the red lines, and CW models by the blue lines.}
    \label{fig:ssfrbar}
    \end{figure}

 \begin{figure}
\center 
	\includegraphics[width=0.7\textwidth]{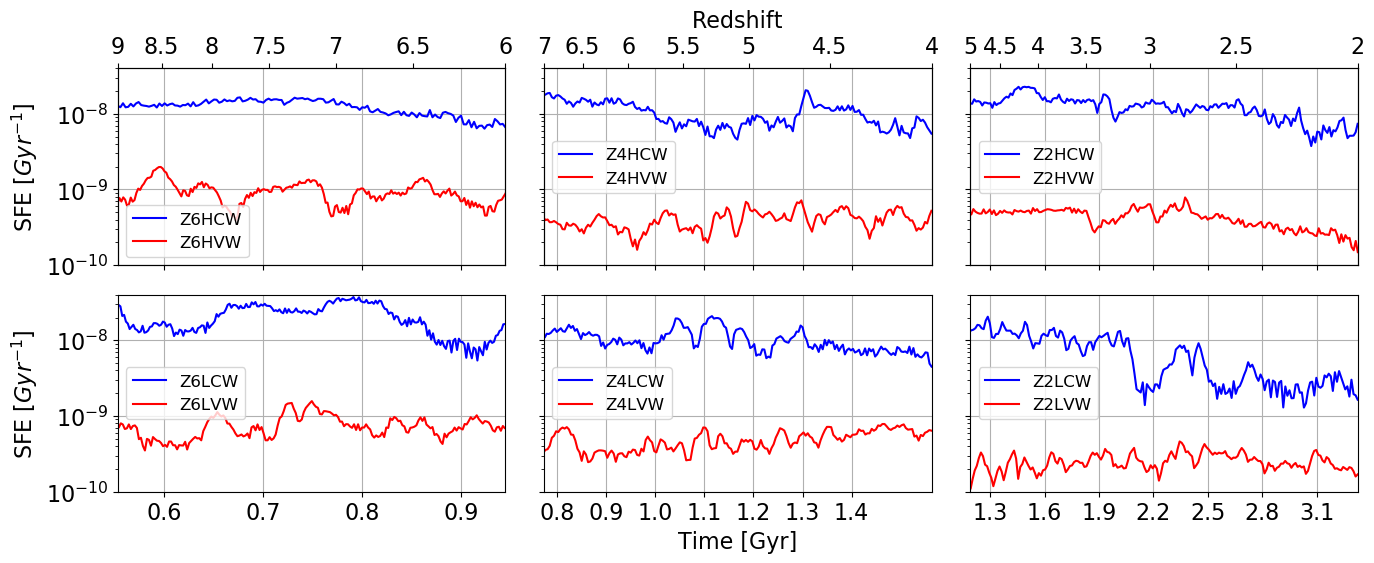}
    \caption{Evolution of star formation efficiency (SFE) in the bar region, defined as the SFR normalized by the gas mass in the bar.  The VW models are given by the red lines, and CW models by the blue lines. }
    \label{fig:SFEbar}
    \end{figure}

 \begin{figure}
\center 
	\includegraphics[width=0.7\textwidth]{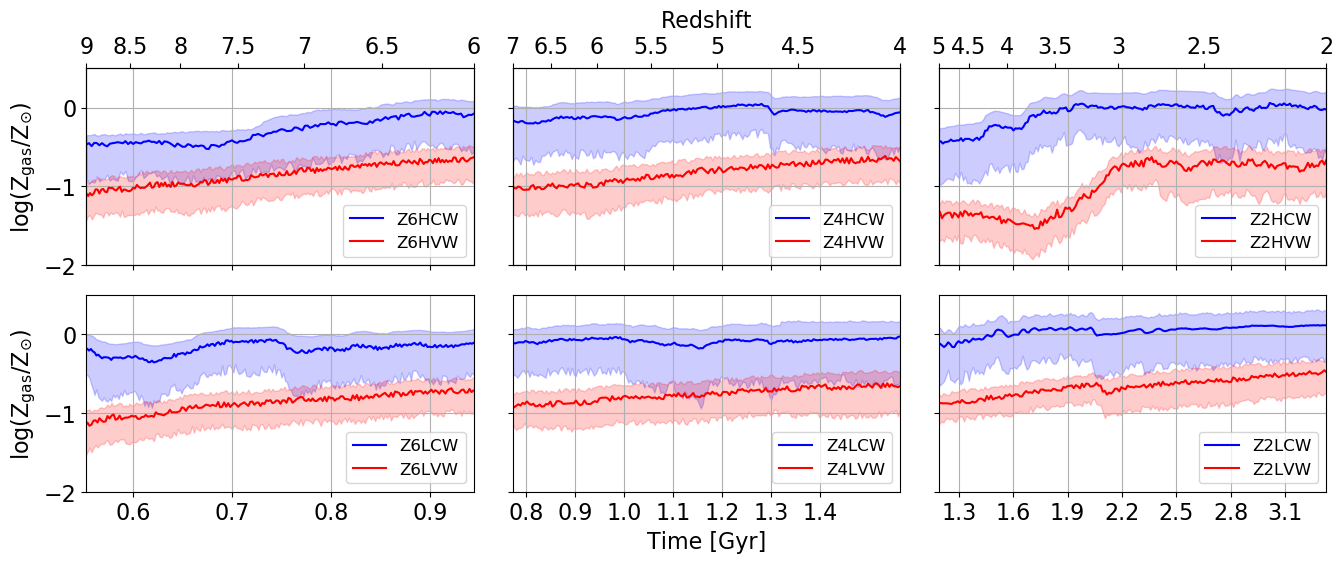}
    \caption{Evolution of gas metallicity in the bars. Note that we find no real difference between the metallicity of gas and stars, both in the bar and in the host galaxy. The shaded area represent the 20\% to 80\% percentiles. The VW models are given by the red lines, and CW models by the blue lines.}
    \label{fig:bargasZ}
    \end{figure}

The SFR within a bar is closely related to the galactic gas reservoir, an additional gas supply from outside, and to the specific gas kinematics (e.g., shear) within the bar. To calculate the SFR inside the bar, we count the stars formed per fixed time period within the bar volume at that time. Of course, stars can move out of this volume with time. 

Figure\,\ref{fig:sfrbar} displays both the SFR in the bars and their parent galaxies. We observe that the SFR in galaxies is typically larger than that in the associated bars, especially in CW galaxies and closer to $z_{\rm f}$. Bursts of star formation in a bar are typically associated with bursts of star formation in the galaxy, but not always, e.g., in Z6HCW around $z\sim 6.4$. The bursts of SF dominate the process in bars much more than in galaxies, where these burst are diluted by the quiescent star formation.

The SFR in galaxies shows a clear evolutionary trend with decreasing $z_{\rm f}$. The $z_{\rm f} = 6$ models exhibit a substantially higher SFR for the CW models, which results in the lower gas fractions in these objects, e.g., Figure\,\ref{fig:gasbar}. The SFRs in bars are similar in CW and VW models, except following major mergers. The $z_{\rm f}=4$ models, the galaxy SFR behaves similarly. The CW bars experience a sharp increase in SFR in tandem with their parent galaxies after a substantial increase in cold accretion influx at $z\sim 4.7$ and 5.7, accordingly. In comparison, in  $z_{\rm f}=2$ models, the SFR is nearly identical in CW and VW bars, but is larger in Z2LCW galaxy.

The specific star formation rate inside the bars, sSFR, i.e., the SFR per unit mass, is shown in Figure\,\ref{fig:ssfrbar}.  We observe a steady decline in sSFR in each model towards $z_{\rm f}$, as well as a much weaker decrease from $z_{\rm f}=6$, to 4 and 2. This weakening is a direct corollary of the imposed requirement that halos masses are independent of $z_{\rm f}$ and the galaxy masses are, therefore, very similar. After a sharp decline at high $z$, the sSFR level off. The small difference between the CW and VW galaxies is related to the stellar feedback. Looking carefully into the source of variability in sSFR, we find that it is the change in the bar sizes and their strength with time due to interactions and cold accretion that are responsible for this trend.

Next, we calculate the efficiency of star formation in the bars with respect to the {\it gas} mass in the bar region, SFE.
Figure\,\ref{fig:SFEbar} shows the efficiency of SFR in this definition in the bars. We have compared the SFE in the bars with those in the parent galaxies and found that the SFE in the bar region is a factor of 2--3 higher during the entire run. We also find that the SFE for CW models is substantially higher than the for the VW models. This applies equally to the bar and to the entire galaxy.

\subsubsection{Metallicity evolution}
\label{sec:metal}

Chemical evolution of our models exhibits a mild monotonic increase in metal abundance within the bars, both in gas (Fig.\,\ref{fig:bargasZ}) and in stars, and most of the time. We have compared the metallicity of the bar region with that of the galaxy outside the bar  and found that the bars are typically more metal-rich by $\sim 30\%$ for $z_{\rm f} = 6$ galaxies, but this difference becomes smaller for later $z_{\rm f}$. Moreover, we do not observe any substantial differences between the evolution of gas and stellar metallicities, and hence, only show the former. We do find that the metallicities have been modified substantially by efficient mixing maintained by the mergers and flybys, as we discuss below. On the other hand, the cold accretion influxes regenerated the metallicity gradients between the bars and the outer disks. Lastly, the CW bars (and galaxies) show larger metallicity compared to VW objects, by a factor of $\sim 5-10$, because of the high SFR in the former.

In some cases. we can directly trace the metallicity evolution to mergers and other violent events. For example, Figure\,\ref{fig:bargasZ}, for model Z2HVW, shows a substantial increase of metallicity in the gas of the bar region by a factor of $\sim 10$, during $z \sim 3.7-3$. This corresponds to the time period of $\sim 0.5$\,Gyr.

Following the evolution of this object, we have observed that during this time it has experienced two major mergers with a metal-rich galaxy, in both stellar and gaseous components. The first merger happens at $z\sim 4.2-3.7$ and the second at $\sim 3.45-3.25$ (Fig.\,\ref{fig:starbarmassfrac}). The steady increase in metallicity is the result of the mixing process between these objects which differ in mass by not more than a factor of two. In fact, the first merger triggers the bar, whose amplitude increases quickly from that of the mild oval distortion to the full-fledged bar, while the second merger strengthens it (Fig.\,\ref{fig:a2}). We also observe a large variation of the stellar bar-to-galaxy mass ratio, $M_{{\rm *bar}}/M_{{\rm *gal}} \sim 0.5-0.8$, during this time period (Fig.\,\ref{fig:starbarmassfrac}).

\subsubsection{Bar triggering}
\label{sec:trig}

As we stated in the Introduction, the bar origin is presently unknown. Among the bar triggering mechanisms listed there, we have detected all with the exception of the bar instability in axisymmetric disks. Among the most prominent triggering and destruction mechanisms in our simulations, we find the asymmetric gas inflow (e.g., Fig.\,\ref{fig:example}a), galaxy interactions (i.e., mergers and flybys, Fig.\,\ref{fig:example}bc), and action of the DM substructure.

Prograde interactions (mergers and flybys), especially those close to the disk plane, are efficient in triggering bars (e.g., , as well as increasing their amplitudes, while retrograde interactions are known to have an opposite effect, e.g., Figure\,\ref{fig:example}c, in accordance with the previous works \citep[e.g.,][]{ger90,bere04,roma08}. In our runs, these effects of prograde and retrograde mergers and flybys have been observed to strengthen and weaken the bars, sometimes to the point of $A_2$ dropping below 0.15, thus forming oval distortions. Filamentary gas accretion onto the outer disk interacts with the bars --- the gas is moving in and being repulsed by the action of the bar torque outside the CR. This leads to enhancement of the outer stellar ring in galaxies forming by the action of the Outer Lindblad Resonance (OLR), especially in a more gas-rich VW galaxies (e.g., Z6HVW and Z4HVW in Figure\,8 of Paper\,I).

\section{Discussion} \label{sec:discussion}
In this work we focus on evolution of stellar bars in high-$z$ galaxies by means of high-resolution zoom-in cosmological simulations and compare them to low-$z$ counterparts. We focus on the main galaxies of similar mass DM halos, terminating their evolution at redshifts $z_{\rm f} = 6$, 4 and 2 and analyze the bar evolution at $z\sim 9-2$. We choose DM halos situated in high and low density environments and implement two different galactic wind feedback mechanisms for each model, namely, the Constant Wind (CW) and the Variable Wind (VW) (section\,\ref{sec:winds}). Further details on properties of simulated galaxies and their DM halos, including their morphologies, have been analyzed and presented in Paper\,I. The baryonic properties of host DM halos and their interactions with the cold accretion flows are further analyzed in Bi et al. (in preparation). Our results of high-$z$ bar evolution have been presented in section\,\ref{sec:results}, and its comparison with the isolated models and those outside the galaxy clusters heavily relies on numerous published works over the last three decades. As a next step we summarize and analyze them. 

\begin{itemize}

\item First and most important is that evolution of stellar bars at high redshifts differs profoundly from from that at low $z$. These differences originate from a much higher gas fraction, larger average mass density in the DM and baryons, from much more active environment associated with frequent mergers and close flybys, and with higher rates of the cold accretion. This leads not only to quantitative but also to qualitative changes in the bar evolution.
    
\item We find that the the bar pattern speeds in all models experience a substantial variability, in a sharp contrast with the low redshift bars which show a monotonic decline in $\Omega_{\rm bar}$. This behavior has its origin both in the internal and external factors. Internally, a weaker feedback in CW models results in a more clumpy galactic disks, and larger SFR. Minor and intermediate mergers as well as close flybys amplify this effect. Moreover, gas inflows being nonuniform in mass and angular momentum speed up and down the bars by increasing the central mass concentration and pumping and removing their angular momentum. This is related to the bars tumbling angular momentum and to internal circulation within the bars. 

The immediate consequence to bar pattern speed variability is the CR-to-bar size ratio, $R_{\rm CR}/R_{\rm bar}$, being not limited to a narrow range of $1.2\pm 0.2$. While we have discussed known cases when this ratio is found outside this range for short time periods, the current modeling shows that one should not expect it to lie within this range at all at high redshifts. This means that one should not expect  the offset dust lanes associated with stellar bars at low $z$ to be present at high $z$ --- an important detail for future observations of high-$z$ galaxies, when these will be available with a sufficient resolution. We note that while low-$z$ galaxies residing in denser environment experience the above interactions these interactions are well separated in time, allowing the system to relax and to resume its quiescent evolution.

\item Third, the bar strength shows a similar erratic behavior not correlated with the bar pattern speed, unlike a clear anti-correlation between bar strength and their pattern speeds shown in isolated models \citep[e.g.,][]{atha03,villa09}. We find no correlation between these characteristic bar properties, i.e., $A_2$ and $\Omega_{\rm bar}$. The origin of this correlation in isolated models lies in the loss of angular momentum by the bar, both in tumbling of its figure and in internal circulation --- the former leads to the bar slowdown and the latter to bar strengthening, and vice versa. Apparently, such processes do not act in tandem when multiple factors affect the bar dynamics. 

While one can approximate galaxies as following the quiescent evolution for prolonged periods of time at low redshifts, it seems an unreasonable approach at high redshifts. The intense cold accretion flow encountered in our models, and frequent mergers and close flybys have no counterparts at low redshifts. As a result, the bar evolution is far from being monotonic at high $z$. This includes basically all the bar properties, i.e., pattern speeds, strength, size and shape, etc. 

\item Fourth, bars at low redshifts are long-lived. Basically, to destroy them, one is required to destroy the parent galactic disks \citep[e.g.,][]{bere04}, or to spin up the parent DM halos \citep{long14,coll18}. This robustness of bars is not maintained at high redshifts in our simulations. We find that not only the high amplitude perturbations by mergers and flybys, but also variable cold accretion inflow onto the disk and its subsequent channeling into the bar region can eliminate the bar by driving its amplitude below the threshold of $A_2/A_0\sim 0.15$ and converting it at best to a very mild oval or completely washing it out. 

Wild swings in the bar amplitude have been observed in our simulations as typical. In addition to abrupt changes in the bar amplitude on a galaxy rotation timescale, we do see a gradual decline or increase in the amplitude due to a weaker variability which originates in internal gas flows within the disk. This evolution displays not only multiple destruction events of stellar bars but also their reformations. 

The galaxies modeled here are sufficiently massive during the time under consideration in this work, $M_*\gtorder 10^{10}\,{\rm M_\odot}$, and embedded in DM halos which are smaller than that of the Milky Way only by a factor of 2.5. They are typical for the contemporary universe, but become progressively rare for higher redshifts considered here. Stellar bars developing in our models also constitute a healthy fraction of the parent galaxies' stellar masses (Fig.\,\ref{fig:starbarmassfrac}). The typical mass fraction of stellar bars in the local universe is $\sim 0.15$ \citep[e.g.,][]{gad09}. In all of our models, the bar mass fraction is $\gtorder 0.3$, especially in the VW models. This is despite that these bars are dominated by the gas. We can compare this result with bars in isolated models, where bars can reach $\sim 0.8$ fraction of the disk mass in $\sim 7$\,Gyr \citep[e.g.,][]{villa09}. A simple conclusion follows from observations --- the bar growth in the universe is either limited in time or the bar-to-disk size ratio is larger than typically obtained in isolated models --- we return to this issue in section\,\ref{sec:num_ratio}. 

Although the high-$z$ bars being more massive, their evolution differs profoundly from that of bars in the contemporary universe, and their resilience at low $z$ is undoubtedly associated with rare merging events and low rate or absent cold accretion flows.

\item Fifth, it would be difficult if not impossible to find bars and galaxies in a contemporary universe with the gas fractions of $\gtorder 50\%$. Yet, all of our models have been found in this regime at least part of the time under consideration here.  Models with a stronger feedback, i.e., VW, display less efficient SF and, therefore, larger gas fractions, up to 80\%. Weaker feedback has led up to $\sim 50\%$ gas in galaxies, and 5\%-50\% in bars. Gas is known to substantially affect the bar properties, e.g., shapes and strength \citep[e.g.,][]{bere98,villa10}. The dissipative character of gas leads to its being channeled inwards and accumulating in the galactic centers, reducing the gas fraction in bars, and modifying and removing a substantial fraction of stellar orbits aligned with and supporting the bar. The resulting increased central mass concentration has an adverse effect on the bar strength and was extensively studied for isolated galaxies \citep[e.g.,][]{bere98,bere07,villa10}. However, while a larger gas fraction does not acts adversely on the bar formation --- bars form early in our simulations, it does modify such properties as the bar size and their growth secularly, as we discuss below.  

The dynamical effects of gas on stellar bar properties have been investigated in detail by \citet{villa10}, which can be used as a benchmark for isolated models. Beyond some dependency on the numerical resolution, the gas fraction leads to a diverging evolution in bar properties, i.e, in the gas-poor below $\sim 5\%$, and the gas-rich bars above $\sim 12\%$. This difference affects all the bar basic parameters such as the bar strength, central mass concentration, bar vertical buckling amplitude, bar size, etc. Furthermore, for some periods of time the stellar bar pattern speeds monotonically decrease in the gas-poor models, while stay flat or even increase slightly in the gas-rich models. The effect of gas component on the overall bar shape, including the boxy/peanut shape bulges resulting form the vertical buckling instability, have been modeled already in \citet{bere98}.

\item Sixth, at $z\gtorder 2$, galaxies in the mass range modeled here are substantially smaller than at $z=0$. Because the dissipative dynamics does not scale the same way as the collisionless dynamics scales, one can anticipate that the basic parameters which characterize the bar properties at high-$z$ differ as well from their low redshift counterparts. Indeed our bars are small in tandem with their parent galaxies.  This decrease in the characteristic sizes of stellar bars can explain the substantial gradients in chemical composition between the bars and outer disks of galaxies. Mergers and flybys induce the radial mixing in the gaseous and stellar components, but have a smaller effect on the very central regions hosting the bars. Moreover, the pristine gas inflows via filaments also do not affect the central regions and restores these gradients by lowering the metallicty in the galaxy disks.

But small sizes of high-$z$ bars and their appearance in every model provides an additional argument that bars are not contemporary morphological features as argued by \citet{sheth08}, but exist at all redshifts where galactic morphology can be defined \citep{jog04}, and only change their basic characteristics, such as sizes, gas fraction, etc. Most importantly, the bars should be not less abundant during the epoch of mergers, and when the galactic disks are characterized by larger dispersion velocities.  

\end{itemize}

\subsection{Origin of bars: spontaneous vs triggering}
\label{sec:origin}

Our choice of comparing the galaxy evolution within similar mass DM halos at the final redshifts, $z_{\rm f}$ = 6, 4, and 2, means that in physical coordinates this is translated to a decreasing average DM density with decreasing $z_{\rm f}$. This is expected to affect the properties of embedded galaxies, as well as their dynamics, but to what extent? It should also affect the intensity of the cold accretion flows and the merger rates and interactions with flybys.

Hence, the environment emerges as the major factor driving evolution of galactic bars at high redshifts in contrast with the local universe where the bar fraction does not correlate with the environment \citep[e.g.,][]{ague09} --- this conclusion, however, is based on a sample biased to noninteracting galaxies.  Moreover, \citet{mari09} found that at $z=0$, the cluster environment does not strongly affect the bar fraction. This conclusion agrees with \citet{hel07} and \citet{roma08} who argued that one should not expect the difference between various environments because the stellar bars can be triggered by the DM substructure.

In contrast, the high-$z$ evolution of stellar bars presented here was found dominated by the environment but in a very specific sense --- all bars observed in our runs have formed not as a result of the bar instability, but being triggered by mergers, flybys and other external factors. It is difficult at all to talk about galaxies having a quiescent evolution which can lead to the bar instability under these circumstances. 

\subsection{Effects of bars on galactic morphology}
\label{sec:internal}

About 50\% of the edge-on local disks exhibit peanut/boxy shape bulges \citep{luti00}, which routinely accompany the vertical buckling instability of stellar bars. This instability supposed to happen when strengthening bars reach some critical amplitude, which typically happens after about a couple of bar rotations --- the bar amplitude grows exponentially during the bar instability. However, tidally-triggered bars reach the maximal amplitude faster because they form with a finite amplitude already. Note that the peanut/boxy-shaped bulges can form without buckling --- this has been demonstrated by \citet{friedli90} --- such bulges can form as a result of resonant interactions between the motions in the disk plane and vertical oscillations, a secular rather than dynamical phenomenon.  

We do detect such obvious peanut/boxy bulge shapes in our models. However, they appear transient, and are quickly washed out by one or more processes. For example, \citet{bere98}  have investigated the effect of the gas on the modeled bar shapes and found that the characteristic boxiness of the bulge is washed out by the gas fraction of $\gtorder 10\%$. The gas modifies the gravitational potential which affects the stability of the 3-D orbits within the bar. Not only the bulge boxiness largely disappears, but the bar amplitude is weakened as well.  

More difficult is to explain the observation that two galaxies in the local universe are caught in the process of buckling \citep{erwin16}. The explanation requires either that the observed bars are young (i.e., are products of a recent bar instability), as bars buckle immediately after reaching their maximal amplitudes. Or they have been triggered tidally very recently, i.e., in much less than a 1\,Gyr. Additional, and at this point a hypothetical suggestion, is that the buckling can be postponed substantially after the bars reach their maximal amplitude --- so bars which formed few Gyrs ago can buckle today.

Following \citet{elzant02} analysis of interactions between a triaxial halo and a stellar bar and associated interactions between double bars \citep{elzant03}, we looked into variability of bar properties in our models with respect to the angles they make with the prolate axis of the parent halos. While bar parameters vary with time including on a short timescales, we do not observe this correlation in any of the models. We  calculated the halo prolateness\footnote{We define the halo prolateness by $\epsilon_{\rm \Phi}=1-b/a$, and the halo flatness by $f_{\rm \Phi}=1-c/a$ based on ellipse fitting of the isopotential contours, where {\it a}, {\it b} and {\it c} are the halo major, intermediate and minor axes \cite[e.g.,][]{bere06}. The isopotentials have a clear advantage over the isodensity contours, being much less noisy.} for the initial redshifts (i.e., $z=9$, 7, and 5) shown in our Figures. By that time, the halo prolateness appears to be washed out within the inner $\sim 20$\,kpc (in physical coordinates), in agreement with baryon action \citep{bere06}. While at higher redshifts halos do expect to be triaxial, the baryon infall removes their prolateness at smaller radii quite efficiently. We calculated that halo prolateness increases with radius outside 20\,kpc, reaches maximum and decreases further out. We attribute this decrease to environmental effects, i.e., mergers and flybys, which  randomize the DM orbits outside 50\,kpc at redshifts modeled here. This process has washed out prolateness in the inner halo, and in tandem with small bars at high redshifts explains why we do not detect the quadrupole interactions between the parent halos and the bars. Of course, this interaction may still exist but has a small amplitude which we did not detect.
 
We should also emphasize additional effects of stellar bars on galactic morphology. One of them is the stellar and gaseous mass redistribution which is associated with angular momentum flows from the bar region to the outer disk and DM halo. In Paper\,I, we have shown the specific imprints of bar dynamics on galactic morphology (see Figure\,6 there). Because the VW bars appear stronger compared to the CW bars over most of the evolution time and larger in size, one can observe formation of the dark rings at about the CR radii and the bright rings just outside them. The origin of these rings follows directly from winding up the stellar/gaseous spiral arms driven by the bar. As they wind up, they form a ring around the OLR, which causes enhanced star formation there. We have noticed but did not analyze this further, that accretion flows from larger radii in galactic disks are terminated around the OLR --- an effect which further enhanced the star formation in the outer rings. Closer to the CR the star formation is depressed as the gas moved outwards by the action of the bar. These rings are clearly delineated in all the VW models. 

\subsection{Comparison with other simulations}
\label{sec:comparison}

The comparison with other cosmological simulations involving stellar bars is difficult because some of the published models analyze bars at lower redshifts, $z\ltorder 1.5$, or consider rather quiescently evolving galaxies which are not subject to mergers and cold accretion flows (section\,\ref{sec:intro}). Their bar pattern speeds and strength evolve exactly as in "classical" stellar bars of isolated galaxies. Most of these works but not all avoid the issues of the bar triggering. These models can reflect the conditions at lower redshifts --- much less frequent mergers, substantially reduced cold accretion flows, much lower gas fraction in the host galaxies, and relaxed DM halos.
  
Focusing on a small number of galaxies but at high resolution provides some advantages. More detailed cosmological simulations of stellar bar evolution have been focused on such quiescently evolving galaxies, in order to mimic the Milky Way galaxy. \citet{oka15} have studied the evolution of stellar bars in two galaxies selected from the Aquarius project \citep{spri08} during $z\sim 1.5-0$. The mass resolution of these simulations has approached the lower end of the isolated models at $z=0$. The spatial resolution was fixed at $\sim 0.26$\,kpc for $z < 3$. The resulting basic parameters of the bars have been shown at $z=1$, 0.5 and 0 only, exhibiting wild oscillations explained as interactions with the stellar clumps. Additional periodic oscillations were explained as interactions with the parent triaxial halos, but the amplitude of these oscillations was high, unlike published so far \citep[][]{bere06,beren06, athan13}. A number of Milky Way-type galaxies have been simulated by the Auriga project \citep{grand17}, with a resolution of 369\,pc at $z=1$. But these bars have been analyzed only at $z=0$ \citep{bla20}.

The main difficulty in comparison of \citet{oka15} work with our simulations is a two-fold: although they discuss the cosmological evolution, all the effects determining the bar evolution appeared to be intrinsic to the disk-halo system, so reflecting an isolated galaxy evolution. For example, the bar pattern speed has been monotonically reduced since $z\sim 1.5$. This is explained as the angular momentum loss to the DM halos. Oscillations in the pattern speed have been attributed to interactions with a triaxial halo. This indeed can happen \citep[e.g.,][]{bere06}. But the effects of mergers and flybys, as well as the cold accretion flows were not detectable, and not discussed or analyzed. 

\citet{scan12} have studied the evolution of two Milky Way-mass galaxies which are mildly isolated. Only the final products have been discussed, at $z=0$. So no details of the bars origin, and evolution of their dynamical properties could be found. Furthermore, \citet{zana19} have performed  simulations of two Milky Way-mass galaxies in the cosmological context during $z\sim 3-0$, but only $z<1$ has been discussed, with the goal of analyzing the effect of stellar feedback on this evolution. The galaxies have been taken from the Eris suite of simulations \citep{gue11}, and followed a quiet merger history (no major mergers after $z=3$). For $z\ltorder 9$, the gravitational softening has been fixed at 0.12\,kpc. The two galaxies have differed also by an additional subgrid physics. 

In \citet{zana19}, the two galaxies analyzed have developed stellar bars at $z = 1.14$ and $z= 0.74$. But these lower redshifts formation of the bars appear to follow from numerical resolution. For extended time intervals, bars strength appeared very low, below the value it is considered being a bar. The bar strength has increased at $z < 1$, still remaining low.  The length of the bar appeared around 1\,kpc in one model, being resolved by about 8-9 softening lengths. So, the limit adopted for the bar size of $>1$\,kpc is related to the spatial resolution in the simulation. The bar pattern speed has shown a monotonic decline with time, apparently due to the absence of the major mergers by construction. The resulting evolution of these bars can be thought as a natural extention of evolution at higher redshifts. 

The zoom-in simulations, such as used in the present work, have the advantage of achieving a high mass and spatial resolution, which are necessary in order to reproduce correctly the stellar and gas dynamics of their hosts. However, they cannot deal with a large number of galaxies. An alternative way to analyze the properties of large galaxy population is to use computational boxes of comoving $50 - 100$\,Mpc and even larger. The tradeoff is of course the resolution. Very promising examples are the Illustris \citep{vogel14} and IllustrisTNG \citep{nel18} simulations which evolve $\sim 10^3-10^4$ objects. Some of the recent works analyze these simulations to study the barred galaxies at higher redshifts. Most of the conclusions aim at such issues as the fraction of barred galaxies and similar statistical properties evolving with time, which are outside the scope of this work. However, two issues, i.e., the bar triggering and their survival over long time periods are closely related to bar evolution discussed here.

The EAGLE simulations have a similar mass and spatial resolutions to Illustris, and detect bar formation only after $z\sim 1.3$ \citep{algor17} (but do not mention what triggers it), which can be related to the limited resolution, e.g., 0.7\,kpc at $z=2.8$. Their strong bars develop in about 10 disk rotations, which take few Gyr. Such a prolonged quiescent time evolution would be incompatible high-$z$ universe analyzed in the present work. This is consistent with our conclusion that at $z>2$ it is difficult to expect such prolonged periods of time which allow for the bar instability to develop. Interestingly, they obtain the corotation-to-bar size ratio of $\sim 10$ at $z=0$, even exceeding our ratio for the high-$z$ bars.

Numerical bars in isolated systems are known to be resilient and long-lived, except under specific conditions. It is known that galaxy interactions can weaken a bar \citep[e.g.,][]{ger90,bere04}, a large gas fraction in the disk and a parent halo triaxiality has a negative effect on its strength for a number of reasons \citep[e.g.,][]{bere98,bere07,villa10,athan13,elzant02}, and, finally, bars weaken dramatically inside fast spinning DM halos \citep[e.g.,][]{long14, coll18}. However, these changes happen rather abruptly on the timescale not exceeding few rotations, and not secularly. 

It is interesting, therefore, that stellar bars in Illustris-1 simulations have been found to weaken secularly, without any sign of interactions, and moreover, the fraction of barred galaxies decreases with redshift to a very low 21\% at $z=0$ as well \citep{pes19}. This analysis relates to objects with $\gtorder 10^5$ stars and hence the stellar galaxy mass of $\gtorder 10^{11}\,M_\odot$. Therefore, the mass resolution is not an issue. However, the spatial resolution is 0.7\,kpc, which for smaller bars is clearly insufficient --- the average bar size in simulation is 4.7\,kpc at $z=0$ and apparently even smaller at higher $z$. In any case, referring to our simulated galaxies, the bar dissolution to an oval distortion could be always traced to one of the known cases listed above.

Bar triggering in Illustris-1 has been attributed to mergers and interactions and only $\sim 7\%$ to the bar instability \citep{pes19}. This conclusion agrees with our analysis, as we could not refer to a single case of a bar instability at $z\gtorder 2$. We found multiple cases when the bar has been dissolved and reformed, accounting for prograde and retrograde mergers and flybys. 

The more advanced TNG100-1 simulations have been capable of increasing the bar fraction to 55\% at $z=0$ \citep{zhao20}. It was found that the bar formation is supressed for $M_* < 10^{10.6}\,M_\odot$, arguing that this is the result of insufficient resolution for bars of a radius of $< 1.4$\,kpc. This indeed appears to be the case, as the spatial resolution at $z\ltorder 1$ is 0.7\,kpc, leading to only two softening lengths per bar radius. However, even larger bars should be affected. Comparing with the stellar bars in our zoom-in simulations, which all are of a sub-kpc size at $z\gtorder 2$, a much higher spatial resolution is required, which is still unattainable in the full box simulations. 

Finally, \citet{rosa20} have used the TNG50 simulations to analyze the bar population for $z\ltorder 4$, limiting it to $M_*\gtorder 10^{10}\,{\rm M_\odot}$ and dominant disks, i.e., $D/T\gtorder 0.5$, which substantially decreases the number of such galaxies. Kinematic decomposition has been used to separate the disks, using \citet{abadi03} and \citet{mari14} method. We compare this method with ours in Paper\,I. We note that our VW galaxies appear just below this limit even at $z_{\rm f} = 6$, 4 and 2. 

Interestingly, these authors separate the barred galaxies at $z\gtorder 2$ from their low redshift counterparts, due to the small numbers at high $z$. For example, at $z=4$, there are only 2 barred galaxies, and at $z=3$, there are 10.  Hence, only bar fraction evolution with redshift = 4, 2, 1, 0.5 and 0 are shown. And the intrinsic properties of bars are shown only at these $z$, which are very similar to those shown here. 

\subsubsection{CR-to-bar size ratio}
\label{sec:num_ratio}

We find that the $R_{\rm CR}/R_{\rm bar}$ ratio is highly variable and often exceeds substantially the canonical range of $1.2\pm 0.2$ (section\,\ref{sec:locals}). The meaning of high $R_{\rm CR}/R_{\rm bar}$ ratio is that bars appear much shorter than their CR radius (i.e., they are slow bars) and that the conditions for the appearance of the offset dust lanes within these bars at low redshifts are not maintained at high $z$. Note that all our galaxy models are very gas-rich, and under these conditions the bar size is shortened as the bars have difficulty to grow in size as they slowdown braking against the DM \citep{villa10}. 

Slow bars have been recently obtained based on the analysis of the IllustrisTNG \citep{nel18} and EAGLE \citep{schaye15} simulations \citep[][]{roshan21} for $z=0$. They preferred an explanation that slow bars in these simulations are the result of excessive braking of the bars against the DM, thus questioning the existence of DM on galactic scales. Such interpretation does not appear warranted based on our results, but may require a more detailed analysis which is outside the scope of this paper.  

\section{Conclusions} \label{sec:conclusions}

In summary, we have analyzed the evolution of stellar bars in host galaxies, which lie in similar mass DM halos of log\,$M_{\rm vir}\sim 11.6\,{\rm M_\odot}$ at $z_{\rm f} = 6$, 4 and 2. We apply two different galactic wind models to assess the effect of weak and strong feedback on the bar and galaxy properties, as well as low and high overdensities of their host halo environments. We find that all modeled galaxies develop stellar bars which are gas-rich. We find that the bar evolution at $z\gtorder 2$ differs dramatically from quiescent evolution of low redshift bars, and even more from bar evolution in isolated halos. Most interesting is the almost complete absence of a monotonic evolution of various parameters characterizing the bars, such as their amplitude and pattern speed, as well as expected high gas fractions within the parent galaxies and their bar regions. This is a direct consequence of ongoing intense buildup of the host galaxies, which includes mergers and close flybys, cold accretion flows and interactions with the DM substructure.  

Bar triggering, weakening and dissolution is a corollary of these processes to the extent that galactic disks lack the ability to axisymmetrisize in the non-axisymmetric background potential. Tidal bar triggering, and not the global bar instability, is a direct consequence of the conditions prevailing during this epoch. Repeated triggering of stellar bars and their frequent dissolution is a signature of this evolution which is expected to subside at lower redshifts.

\section*{Acknowledgements} 
We thank Phil Hopkins for providing us with the latest version of GIZMO.
We are grateful to Alessandro Lupi for his help with GIZMO. We also thank Xingchen Li for sharing some of the analysis software. I.S. is grateful for a generous support from the International Joint Research Promotion Program at Osaka University. This work has also been partially supported by the Hubble Theory grant HST-AR-14584, and by JSPS KAKENHI grant 16H02163 (to I.S.).  The STScI is operated by the AURA, Inc., under NASA contract NAS5-26555. E.R.D. acknowledges support of the Collaborative Research Center 956, subproject C4, funded by the Deutsche Forschungsgemeinschaft (DFG).  Simulations have been performed using generous allocation of computing time on the XSEDE machines under the NSF grant TG-AST190016, and by the University of Kentucky Lipscomb Computing Cluster. We thank Vikram Gazula at the Center for Computational Studies at the University of Kentucky for his continuous help.

\section*{Data Availability}

The data presented in this work can be obtained upon request.









\bibliographystyle{aasjournal}
{}



\end{document}